\documentclass[twocolumn,longauthor]{aastex62}
\usepackage{apjfonts}
\usepackage{amssymb, amsmath}
\usepackage{color}
\usepackage{natbib}

\shorttitle{Around the Way}
\shortauthors{Dai, Robertson, \& Madau}

\providecommand{\adsurl}[1]{\href{#1}{ADS}}
\newcommand{\LCDM}{\Lambda\mathrm{CDM}}
\newcommand{\Msun}{M_{\odot}}
\newcommand{\pc}{\mathrm{pc}}
\newcommand{\kpc}{\mathrm{kpc}}
\newcommand{\Mpc}{\mathrm{Mpc}}
\newcommand{\vir}{\mathrm{vir}}
\newcommand{\kms}{\mathrm{km}\,\mathrm{s}^{-1}}
\newcommand{\masyr}{\mathrm{mas}\,\mathrm{yr}^{-1}}
\newcommand{\muasyr}{\mu\mathrm{as}\,\mathrm{yr}^{-1}}

\newcommand{\Myr}{\mathrm{Myr}}
\newcommand{\Gyr}{\mathrm{Gyr}}

\newcommand{\qz}{q_z}
\newcommand{\qzh}{q_{z,h}}

\newcommand{\Mvir}{M_\vir}

\begin{document}

\title{Around The Way: Testing $\Lambda$CDM with Milky Way Stellar Stream Constraints}


\author{Biwei Dai}
\affiliation{Department of Physics, Peking University, Beijing, 100871, China}
\affiliation{Department of Astronomy \& Astrophysics, University of California, Santa Cruz, 1156 High Street, Santa Cruz, CA 95064, USA}

\author[0000-0002-4271-0364]{Brant E. Robertson}
\affiliation{Department of Astronomy \& Astrophysics, University of California, Santa Cruz, 1156 High Street, Santa Cruz, CA 95064, USA}

\author{Piero~Madau}
\affiliation{Department of Astronomy \& Astrophysics, University of California, Santa Cruz, 1156 High Street, Santa Cruz, CA 95064, USA}
\affiliation{Institut d'Astrophysique de Paris, Sorbonne Universit{\'e}s, UPMC Univ Paris 6 et CNRS, UMR 7095, 98 bis bd Arago, 75014 Paris, France}

\begin{abstract}
Recent analyses of the Pal 5 and GD-1 tidal streams suggest that the inner dark matter halo of the Milky Way is close to spherical, in tension with predictions from collisionless N-body simulations of cosmological structure formation. 
We use the {\it Eris} simulation to test whether the combination of dissipative physics and hierarchical structure formation can produce Milky Way-like galaxies whose dark matter halos match the tidal stream constraints from the GD-1 and Pal 5 clusters. We use a dynamical model of
the simulated {\it Eris} galaxy to generate many realizations of the GD-1 and Pal 5 tidal streams, marginalize over observational uncertainties in the cluster galactocentric positions and velocities, and compare with the observational constraints. We find that the total density and 
potential of {\it Eris} contributed by baryons and dark matter satisfies constraints from the existing Milky Way stellar stream data, 
as the baryons both round and redistribute the dark matter during the dissipative formation of the galaxy, and provide a centrally-concentrated mass distribution that rounds
the inner potential. The {\it Eris} dark matter halo or a spherical Navarro-Frenk-White dark matter work comparably well in modeling the stream data.
In contrast, the equivalent dark matter-only {\it ErisDark} simulation produces a prolate halo that cannot reproduce the observed stream
data. The on-going {\it Gaia} mission will provide decisive tests of the consistency between $\LCDM$ and Milky Way streams, and should distinguish
between models like {\it Eris} and more spherical halos.
\end{abstract}

\keywords{dark matter ---
          galaxies: fundamental parameters ---
          galaxies: halos ---
          galaxies: kinematics and dynamics ---
          galaxies: spiral ---
          galaxies: structure}

\section{INTRODUCTION}
\label{introduction}

The standard $\LCDM$ model successfully explains a wide range of cosmological 
observations on scales $k\lesssim10 h~\Mpc^{-1}$. 
On smaller scales, comparisons between cosmological N-body simulations and observations 
have identified several seeming inconsistencies including the 
cusp-core \citep[e.g.,][]{flores1994a, moore1994a}, missing satellite \citep{klypin1999a, moore1999a}, and Too Big To Fail \citep{boylan2011a} problems
that involve theoretical predictions for the spherically-averaged radial density profile or
substructure of dark matter halos.
These studies have emphasized the need to consider carefully the results of dissipationless
N-body simulations when comparing with data,
and indeed
recent studies with hydrodynamical simulations suggest that baryonic physics may alleviate
some of the apparent tensions
\citep[e.g.,][]{governato2012a, pontzen2012a, brooks2014a, madau2014a, chan2015a, onorbe2015a, wetzel2016a}.
The shape of the halo represents an important additional prediction of galaxy formation simulations
testable through observations. In this paper, we explore whether cosmological simulations
of Milky Way-like galaxies can produce realistic dark matter halo shapes when
modeling the dissipative effects of baryonic physics.

Collisionless simulations predict triaxial or near-prolate halos, with typical density profile
axis ratios $[b/a,c/a]\sim0.4-0.6$ \citep[e.g.,][]{frenk1988a, dubinski1991a, warren1992a, cole1996a, jing2002a, vera2011a}. For external galaxies, observations utilizing gravitational lensing \citep[e.g.,][]{hoekstra2004a, parker2007a} or X-ray isophotes \citep[e.g.,][]{buote1998a, buote2002a} can test such predictions for halo shapes. 
These studies find typical halo ellipticities of $\bar{e}\sim0.35$, 
in agreement with the results from numerical simulations.
Shape constraints for the Milky Way dark matter halo mostly derive from measurements of the density distribution and kinematics of halo stars or the dynamics of tidal streams such as Sagittarius. Current constraints vary, with studies showing that the Milky Way dark matter halo features an oblate \citep[e.g.,][]{johnston2005a, loebman2014a}, close to spherical \citep[e.g.,][]{fellhauer2006a, smith2009a}, prolate \citep[e.g.,][]{helmi2004a, bowden2016a}, or triaxial \citep{law2010a} shape.

Constraints on the shape of the Galactic dark matter halo derived from either stellar halo properties or the Sagittarius stream face 
complexities from dynamically-hot stellar populations, unknown accretion histories, and
large angular extents on the sky. In contrast, the comparably dynamically-cold
tidal streams of the globular clusters
Pal 5 \citep{odenkirchen2001a} and GD-1 \citep{grillmair2006a} lend themselves
more readily to dynamical modeling, and have yielded constraints on the shape of the
inner halo of the Milky Way \citep{koposov2010a, bowden2015a, kupper2015a, pearson2015a, bovy2016a}.
\citet{bowden2015a} used a streakline method to model GD-1 in a logarithmic potential model and found a total potential minor-to-major
axis ratio of $\qz=0.90^{+0.05}_{-0.10}$. 
\citet{kupper2015a} used the same approach to model Pal 5 in a three-component potential model consisting of a bulge, a disk, and a halo, and derived a halo potential minor-to-major axis ratio of 
$\qzh=0.95^{+0.16}_{-0.12}$. 
 Using the action-angle method of \citet{bovy2014a}, 
 \citet{bovy2016a} modeled the combined data from GD-1 and Pal 5 to determine the axis ratio of the dark matter halo density distribution. Assuming a three-component potential model, they inferred a halo density profile 
 axis ratio of $c/a= 1.05 \pm 0.14$, 
 corresponding to a halo potential flattening close to $\qzh = 1.0$. 
 These results show consistency with each other but appear in tension with the predictions for halo
 shapes from collisionless N-body cosmological simulations. Our work revisits the constraints the Pal 5 and GD-1 stellar streams place on the shape of halo in the context of a self-consistent hydrodynamical cosmological model for the formation of Milky Way-like galaxies, the {\it Eris}
 simulation \citep{guedes2011a}.

Hydrodynamical simulations have long demonstrated that baryonic dissipation
can change the shape of dark matter halos by making them more spherical \citep[e.g.,][]{dubinski1994a, kazantzidis2004a, gustafsson2006a, debattista2008a, abadi2010a, kazantzidis2010a, tissera2010a}.
Dissipation can also redistribute the dark matter mass radially via adiabatic contraction
\citep{blumenthal1986a,gnedin2004a}.
The triaxiality of the halo originates in part from box orbits of the dark matter particles.
The deepened potential well in the center of a galaxy caused by the condensation of baryons can scatter the box orbits and result in a rounder halo \citep{dubinski1994a, debattista2008a}. \citet{kazantzidis2004a},  \citet{gustafsson2006a}, and \citet{abadi2010a} compared the shape of the dark matter halos from collisionless and hydrodynamical simulations,
and found that baryonic dissipation causes halos to grow rounder and mildly oblate in 
their inner regions, with their flattening aligned with any disks formed. 
Baryonic dissipation increased the halo axis ratios by a few tenths, but did not make them spherical. 
It therefore remains unclear whether baryonic dissipation can fully explain the discrepancy between
halo shapes inferred from stellar streams and the predictions of cosmological N-body simulations.
Practical difficulties in forming realistic Milky Way-like galaxies in cosmological hydrodynamical
simulations and in comparing the hydrodynamical simulation results with stream data have limited
previous studies to simplified models for the dark matter halo structure. 

In what follows, we use {\it Eris} as an example model of a Milky Way-like galaxy and simulate the properties
of the Pal 5 and GD-1 streams in this potential.
By marginalizing over the uncertainties in the globular cluster positions
and velocities, we can determine whether the {\it Eris} halo shape permits the observed stream orbits.
In Section \ref{sec:simulation}, we provide an overview of the {\it Eris} simulation, provide details on luminous and dark matter properties of the simulated galaxy, and compare these properties with the observed properties of the Milky Way. 
Section \ref{sec:method} describes our self-consistent field method used to efficiently and accurately represent the gravitational potential of {\it Eris}. 
In Section \ref{sec:shape}, we
measure and compare the shapes of the dark matter halos from {\it Eris} and the equivalent
dark matter-only simulation {\it ErisDark} to study the effects of baryonic dissipation.
Section \ref{sec:tidal streams} describes our method to model and fit the tidal streams of Pal 5 and GD-1 in a variety of potentials, and we present the resulting comparisons with observations in Section \ref{sec:results}. 
We discuss our results in Section \ref{sec:discussion} and present our conclusions in Section \ref{sec:conclusions}.

\section{ERIS SIMULATION}
\label{sec:simulation}

To generate a cosmologically-motivated model for a Milky Way-like potential we use the
results of {\it Eris}, a 
cosmological zoom-in simulation of the formation and evolution of a Milky Way-like galaxy \citep{guedes2011a}.
The simulation evolved a periodic volume $90~\Mpc$ on a side
using the N-body + Smoothed Particle Hydrodynamics code GASOLINE \citep{wadsley2004a}, over the redshift
range $z=90$ to the present in a $\LCDM$ cosmology ($\Omega_M=0.24$, $\Omega_b=0.042$, $H_0=73~\kms~\Mpc^{-1}$, $n=0.96$, $\sigma_8=0.76$). 
The target halo experienced a quiet late merger history and reached a total mass of
$\Mvir=8\times10^{11}~\Msun$, similar to the Milky Way.
The corresponding resampled region contained 13 million high resolution dark matter particles and an equal number of gas particles, for a mass resolution of $m_{DM}=9.8\times10^4~\Msun$, $m_{SPH}=2\times10^4~\Msun$. The simulation used a gravitational softening length of $\epsilon_G=120~\pc$ for all particles, and included Compton cooling, atomic cooling, metallicity-dependent radiative cooling at low temperatures, and heating from a uniform UV background. The star formation algorithm allowed for
the stochastic spawning of
stellar particles in cold, dense regions, each formed with an initial mass of $m_{*}=6\times10^3~\Msun$. The feedback implementation followed \citet{stinson2006a}, with supernova explosions injecting energy and metals in the surrounding gas.

\begin{table*}[tp]
    \renewcommand\arraystretch{1.5}
    \centering
    \caption {\label{table:property} Comparison of the Milky Way and the {\it Eris} simulation}
    \begin{tabular}{ p{7cm}<{\centering} p{7cm}<{\centering} p{3cm}<{\centering}  }
    \hline
    Property & Milky Way & {\it Eris}\\
    \hline
    Dark matter mass $M_{\mathrm{DM}}(<20~\kpc)$ & $1.1\pm0.1\times10^{11}~\Msun$\tablenotemark{a} & $1.1\times10^{11}~\Msun$ \\
    Total mass $M_{\mathrm{tot}}(<60\kpc)$ & $4.0\pm0.7\times10^{11}~\Msun$\tablenotemark{b} & $3.34\times10^{11}~M_{\odot}$\\
    Disk stellar mass $M_{\star, \mathrm{d}}$ & $4.6\pm0.3\times10^{10}~\Msun$\tablenotemark{c}& $3.0\times10^{10}~\Msun$\\
    Disk scale length $R_\mathrm{d}$ & $2.3\pm0.6~\kpc$\tablenotemark{d}& $2.5~\kpc$\\ 
    Bulge stellar mass $M_\mathrm{b}$ & $(0.6-1.6)\times10^{10}~\Msun$\tablenotemark{e,f}& $0.95\times10^{10}~\Msun$\\
    Circular velocity at the Solar circle $V_\mathrm{c}$& $218\pm10~\kms$\tablenotemark{c} & $212~\kms$\\
    Vertical Force at the Solar circle, $\left|z\right|\leq1.1~\kpc$ & $67\pm6(2\pi G \Msun~\pc^{-2})$\tablenotemark{c}& $55(2\pi G \Msun~\pc^{-2})$\\
    \hline
    \end{tabular}
\tablerefs{$^a$\citet{bovy2016a}, $^b$\citet{xue2008a}, $^c$\citet{bovy2013a}, $^d$\citet{hammer2007a}, $^e$\citet{robin2012a}, $^f$\citet{portail2015a}}
\end{table*}

By $z=0$, {\it Eris} displays a late-type spiral morphology and resembles a close Milky Way analog.
Within the {\it Eris} virial radius $r_{vir}\approx235~\kpc$ lie 7.0 million dark matter, 3.0 million gas, and 8.6 million stellar particles. 
Table \ref{table:property} summarizes the basic properties of {\it Eris} and the corresponding observational constraints of the Milky Way. Most simulated properties show consistency with the observations, assuming a solar galactocentric radius of $R_0=8.0~\kpc$. 
The reported measurements of the disk and bulge in the central region follow the orbital circularity-based decomposition
by \citet{guedes2013a}. 
The quantity $\epsilon=j_z/j_c$ describes the orbital circularity of the baryonic (gas and stellar) particles, 
and depends on the angular momentum of each particle along
the $z$-direction $j_z$ and the maximum angular momentum $j_c=\sqrt{GM(<r)r}$. 
In this decomposition,
stellar particles that lie within a radius $r<2~\kpc$ 
and occupy kinematically warm orbits ($\epsilon<0.5$) get assigned to the bulge, while the remaining particles comprise the stellar disk. 
For more details about the {\it Eris} simulation, we refer the reader to \citet{guedes2011a}.

For comparison, we also use an equivalent, dark matter-only simulation called {\it ErisDark} \citep{pillepich2014a}. 
The detailed properties of the dark matter halos in {\it Eris} and {\it ErisDark} are discussed Section \ref{sec:shape}, and a comparison of their properties enable 
us quantify the effects of baryonic dissipation. We refer the reader to
\citet{pillepich2014a} for other analyses of the dark matter halos formed
in the {\it Eris} and {\it ErisDark} simulations.

\section{METHODOLOGY}
\label{sec:method}

In comparing the Pal 5 and GD-1 stellar stream data to the {\it Eris} simulation, we must compute the gravitational forces at the positions of stars in the streams. To marginalize over observational uncertainties in the Pal 5 and GD-1 galactocentric positions and velocities, we perform numerical simulations of the stream formation and measure the resulting likelihood of the observed stream position and velocity data given the {\it Eris} potential model. Since this approach requires evaluating the simulated halo potential, $\Phi$, many tens of billions of times, we need an efficient method
determining $\Phi$ free from the computation burden of using the
simulated particle data. The self-consistent field (SCF) method \citep[e.g.,][]{clutton-brock1973a, hernquist1992a}
allows us to construct an accurate model of the {\it Eris} potential from the particle data using a basis-set expansion.
We then use this SCF model for the {\it Eris} potential to simulate the evolution of the stellar streams in a
manner analogous to the idealized analytical potentials used by, e.g., \citet{pearson2015a} and \citet{bovy2016a}. By including contributions from the baryonic components of {\it Eris},
our approach extends the methodology of \citet{ngan2015a} and \citet{ngan2016a} who used an SCF model to represent the Via Lactea II \citep{diemand2008a} halo potential to compute both tidal debris from accreted systems and tidal streams from disrupted globular clusters.

\subsection{Self-Consistent Field Method}
\label{sec:scf}

The SCF method models dynamical systems using a complete set of bi-orthogonal density-potential pair basis functions
\begin{align}
    \rho(r,\theta,\phi)=\sum_{n,l,m}A_{nlm}{\rho}_{nlm}(r,\theta,\phi) \\
    \Phi(r,\theta,\phi)=\sum_{n,l,m}A_{nlm}{\Phi}_{nlm}(r,\theta,\phi) 
\end{align}
\noindent
where $\rho(r,\theta,\phi)$ represents the density at spherical coordinates $[r,\theta,\phi]$, $\Phi$ is
the corresponding gravitational potential, and $A_{nlm}$ are numerical coefficients that
weight the contribution of the density ($\rho_{nlm}$) and potential ($\Phi_{nlm}$) eigenfunctions
to the expansion.

Representing the density and potential of an arbitrary structure exactly 
requires an infinite number of terms in the expansion. 
In practice one truncates the expansion to remove high frequency terms with
high ``quantum'' numbers $nlm$, resulting in a smoothed potential.
To maximize the efficiency of the expansion, it is advantageous
to select basis functions such that the first few, low-quantum number
terms provide a good approximation to the modeled structure.
Several analytical basis sets have been proposed \citep{clutton-brock1973a, hernquist1992a, zhao1996a, rahmati2009a}, 
but the low-order radial eigenfunctions of these models do not follow the shape
of the Navarro-Frenk-White profile \citep[NFW;][]{navarro1996a, navarro1997a}
with an inner logarithmic slope of $\gamma\sim-1$ transitioning to $\gamma\sim-3$ in the
exterior. Following \citet{weinberg1999a}, we construct a set of 
density-potential pair basis functions for the dark matter halo by solving the
Sturm-Liouville equation numerically. The basis set functions were tabulated
with radius and stored in lookup tables for interpolation.

To compute the basis functions, we assume their form in spherical coordinates follows
\begin{align}
    \Phi_{nlm}(r,\theta,\phi) &=\Phi_{NFW}(r)u_{nl}(r)Y_{lm}(\theta,\phi)\\ 
    \rho_{nlm}(r,\theta,\phi) &=\lambda_{nl}\rho_{NFW}(r)u_{nl}(r)Y_{lm}(\theta,\phi)
\end{align}
where $u_{nl}(r)$ are radial eigenfunctions, $Y_{lm}(\theta,\phi)$ are spherical harmonics, $\lambda_{nl}$ are unknown constants, 
and $\rho_{NFW}(r)$ and $\Phi_{NFW}(r)$ are the NFW density and potential profiles.
Each density-potential pair should satisfy the Poisson equation
\begin{equation}
    \nabla^2\Phi_{nlm}=4\pi G\rho_{nlm}.
\end{equation}
This constraint leads to a standard Sturm-Liouville equation
\begin{equation}
    -\frac{d}{dr}\left[p(r)\frac{du_{nl}}{dr}\right]+s(r)u_{nl}=\lambda w(r)u_{nl},
\end{equation}
where
\begin{align}
    p(r) &= \log^2(1+r)\\
    s(r) &= \frac{\log(1+r)}{(1+r)^2}+\frac{l(l+1)}{r^2}\log^2(1+r)\\
    w(r) &= \frac{\log(1+r)}{(1+r)^2}.
\end{align}
With appropriate boundary conditions, these equations were solved to determine $u_{nl}$ and $\lambda_{nl}$ numerically using Matslise 2.0 \citep{ledoux2016a}. 
The boundary point at $r=0$ is singular, while the outer boundary is set to $u_{nl}'=0$ at $r_{\mathrm{max}} = 32r_s$. 
Since the Milky Way virial concentration ratio is $c_{\vir}\equiv r_{\vir}/r_s\ll32$,
and the region of interest for this calculation (where tidal streams are located) is at galactocentric radii $r\sim r_s$,
placing the outer boundary condition at $32r_s$ has little affect on our results. 
We assume that the potential is symmetric about the $\theta=\pi/2$ plane and azimuthally symmetric. These assumptions allow us to keep only the
even $l$ and $m=0$ terms in the expansion.
This approach provides a fast way to compute the potential from the particle distribution. For
the {\it Eris} simulation, 
this method proves two or three orders of magnitude faster than using a direct summation, depending on which angular symmetries 
for the potential are assumed. 
We verified our method by computing expansions for the
\citet{hernquist1990a} model and
the triaxial NFW model with constant axis ratio
\begin{equation}
\rho(m) = \frac{\rho_s}{m/r_s(1+m/r_s)^2},
\end{equation}
where
\begin{equation}
m^2=x^2+\frac{y^2}{(b/a)^2}+\frac{z^2}{(c/a)^2},
\end{equation}
\noindent 
and $a$, $b$, and $c$ are the major, intermediate, and
minor axis scale lengths.
As expected,
our SCF method works well when modeling triaxial NFW halos but proves inefficient for highly flattened systems
owing to the use of a spherical NFW profile as the lowest-order term.
For modeling the disk component of {\it Eris}, we
include the additional contributions to the potential described below.

\subsection{{\it Eris} Potential Model}
\label{subsec:model}

\begin{figure*}[htp]
\centering
\includegraphics[width=\linewidth, clip]{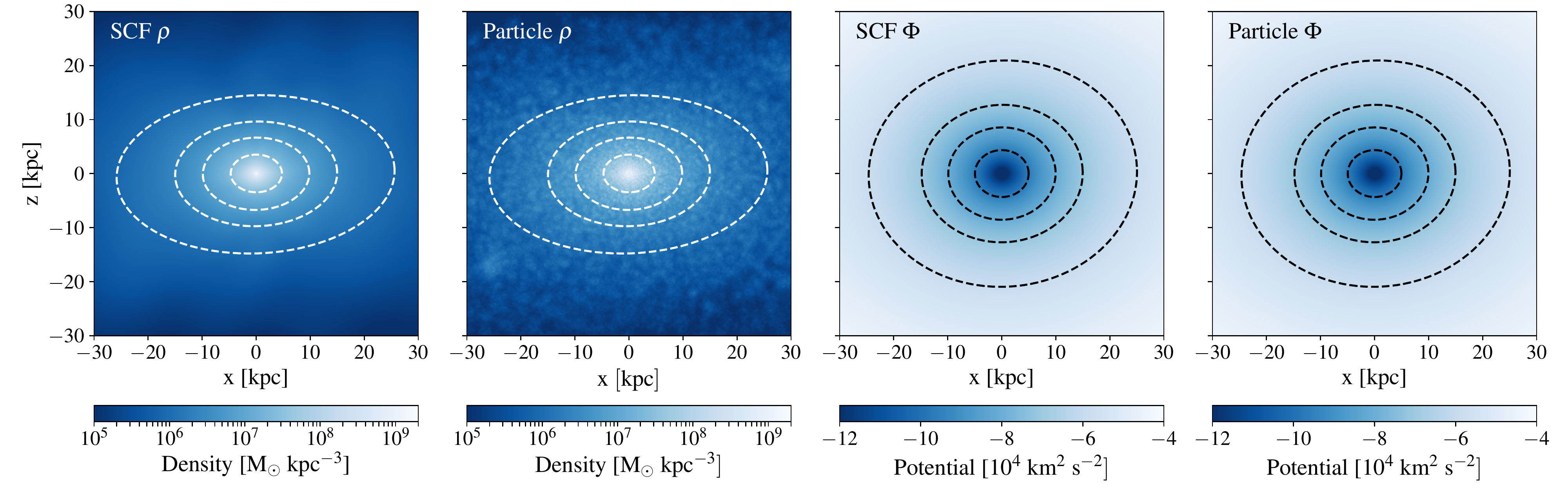}
\caption{\label{fig:halo}
Density (left two panels) and potential (right two panels) fields of the dark matter halo in the {\it Eris} simulation. The first and third panels are calculated from a self-consistent field (SCF) model for the {\it Eris} potential,
while
the second and fourth are measured directly from particles.
The major axis of the dark halo is
aligned with the $x$-axis, and the minor (disk angular momentum) 
axis is aligned with the $z$-axis.
The dashed lines display the best fit elliptical isodensity and isopotential contours calculated from the SCF model.}
\bigskip
\end{figure*}

At least four components are needed to model the complex structure of {\it Eris}: a dark
matter halo, a stellar halo, a bulge, and a baryonic (gaseous and stellar) disk.
To perform our decomposition of the {\it Eris} potential,
we adopt the potential minimum computed by the simulation
as the center of the galaxy.
The total angular momentum vector of all baryonic particles at distances $r<5~\kpc$ from the
center is calculated, and its orientation used to define the $z-$axis of the galactic
disk. We apply a rotation to diagonalize the inertial tensor constructed from the particle
positions in the $x-y$ plane, completing the coordinate system for our expansions. In constructing the separate bulge, halo, and disk expansions for {\it Eris} we aim to
provide a quality representation of the total galactic potential,
and therefore separate the
baryonic particles into distinct structures that approximate the low order terms in
each density profile expansion. We first select baryonic particles in an exponential
disk distribution and fit their density profile with a combination of 
three Miyamoto-Nagai disks \citep{miyamoto1975a, flynn1996a}. 
The remaining baryonic particles are decomposed into a compact bulge and an
extended halo resembling an NFW profile, and then both 
components are modeled using full SCF expansions.
This practical method leads to a ``disk'' vs ``bulge'' decomposition that differs from
the kinematic decompositions reported in 
\citet{guedes2013a} and Table \ref{table:property}, 
but provides an accurate representation of the
total potential.
The dark matter density and potential in {\it Eris}
are separately modeled using a full SCF expansion as described above in 
Section \ref{sec:scf},
and the dark matter halo of {\it ErisDark} is treated in a similar manner. 

Figure \ref{fig:halo} shows a comparison of the density and potential field of the {\it Eris} dark matter halo computed from our SCF expansion (the first and the third panels) and the particle distribution (the second and the fourth panels). In the SCF expansion we include
terms with quantum numbers $n\leq10$ and $l\leq10$, for a total of 1331 terms.
We use direct summation to compute the particle-based potential.
The particle density field was smoothed using a spatially-adaptive polynomial
kernel, with smoothing lengths for each particle computed from the distance
to the nearest 32 neighboring particles.
The smoothness of the SCF density distribution relative to the particle
density field results from truncating the high order terms in the expansion,
but as the isodensity and isopotential
contours illustrate the shape of the halo 
is well-reproduced. Increasing the number of terms in the expansion does not
significantly improve the agreement between the SCF and particle
density and potential fields on angular scales relevant to the stream orbits
we attempt to model.

One concern about the SCF expansion regards the contributions of 
substructures, such as dark matter subhalos,
to the coefficients of the expansion. Although the angular resolution
of the expansion at $l\sim10$ does not resolve nearby subhalos, the particles in
these substructures do contribute power to the expansion coefficients for the $n$ and $l$ terms we retained and
therefore could potentially bias the expansion results. 
For example, if a massive subhalo appears at the major axis of the halo, then the low resolution expansion may give us a more flattened halo. To check the impact of
substructures we compute a test expansion where we 
identify and reduce local density maxima in 
substructures in {\it Eris} simulation, and then recompute the
coefficients on the smoother particle distribution. 
This procedure removes about $8\%$ of the total particles through out the whole halo. We randomly select 1 million points at radii $5~\kpc<r<30~\kpc$, and measure the relative difference of local potential flattening calculated from the two expansions. The local potential flattening is defined as
\begin{equation}
q_{z, h,\rm{local}}=\sqrt{\frac{Z}{R}\frac{F_{R}}{F_Z}},
\end{equation}
which equals the axis ratio of the isopotential contour if the latter is a perfect ellipse. Here $Z$ and $R$ correspond to the vertical and radial
cylindrical polar coordinates, while $F_R$ and $F_Z$ reflect 
the projected components of the gravitational force in those directions.

We find that the presence or absence of substructures does not substantially affect the shape of the potential field. The maximum relative difference of the local potential flattening once substructures are removed is only $0.6\%$. The high order coefficients of the SCF expansion do change after removing the substructures, but these terms have little contribution to the total potential field. The low order coefficients, such as $A_{000}$ and $A_{020}$, are reduced correspondingly because of the loss of mass, but their ratio $A_{020}/A_{000}$
that strongly correlates with the shape of the potential field remains almost unchanged
(the relative difference is $0.3\%$).
We therefore conclude that the presence of substructures do not bias the shape of our halo potential model.
We also consider contributions of power from the Poisson noise associated with
the finite particle number on the expansion coefficients, and for the
high-resolution {\it Eris} simulation the contribution remains small on 
the radial and angular scales retained in our expansion. 

\section{Density and Potential Shape of {\it Eris}}
\label{sec:shape}

\begin{figure*}[htp]
\centering
\includegraphics[width=\linewidth, clip]{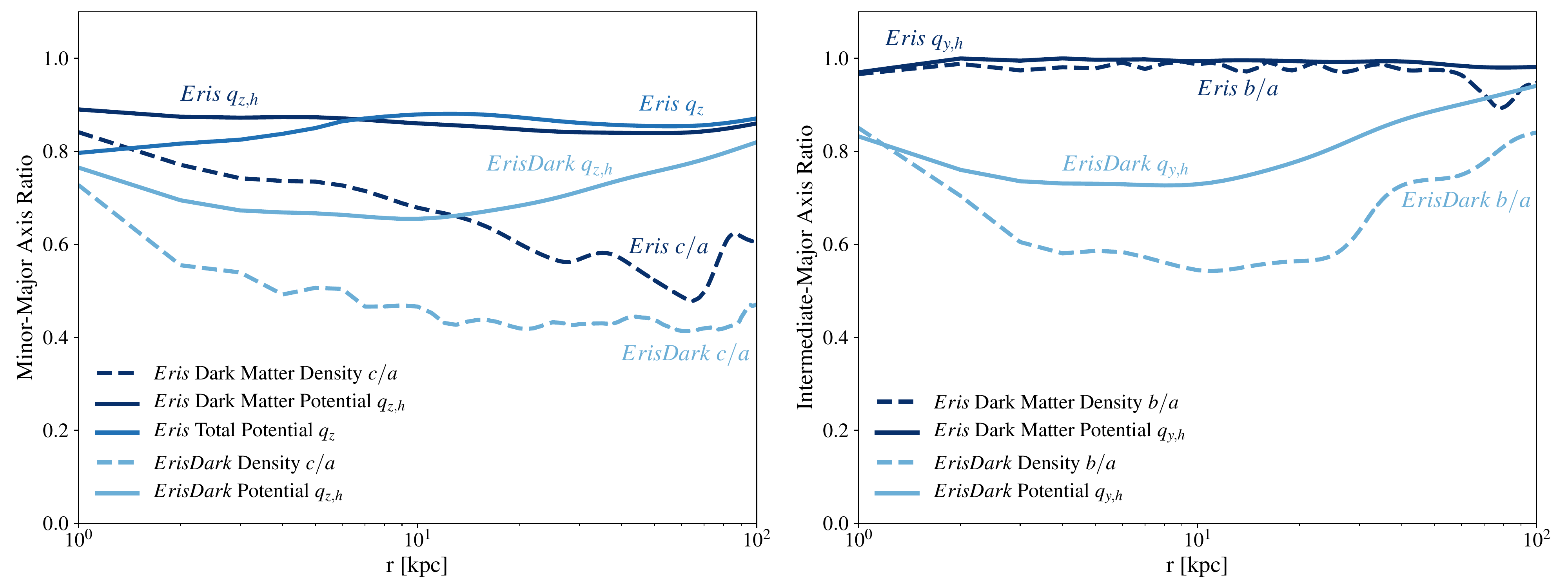}
\caption{Minor-to-major axis ratio (left) and intermediate-to-major axis ratio (right) of our potential models as a function of radius. The dark blue color displays the dark matter halo in {\it Eris} simulation, medium blue shows the total potential in {\it Eris}, and light blue represents {\it ErisDark} simulation. The solid lines show the potential flattening, and the dashed lines
display the shape of the density field.}
\label{fig:shape}
\bigskip
\end{figure*}

Figure \ref{fig:shape} shows the minor/major (left panel) and intermediate/major (right panel) axis ratios for the density and potential in the {\it Eris} (dark blue) and {\it ErisDark} (light blue) simulations. 
We fit the isodensity and isopotential contours with ellipsoids, defining
the major, intermediate and minor axes of the ellipsoids relative to the reference frame formed by the galaxy angular momentum and 
diagonalization of moment of inertia tensor (see Section \ref{sec:method} above).
In the $x-y$ and $x-z$ planes, we fit these contours with ellipses every $\Delta r=1~\kpc$ over radii $r=1-100~\kpc$ with the center [$x_c$, $y_c$], the semi-axes $a$ and $b$, and the tilt angle $\tau$ of ellipses allowed to vary. The best fit $x_c$, $y_c$ and $\tau$ remain close to zero, suggesting that our descriptions for the galaxy centroid and principle axes 
are accurate.

At radii $r\sim10-20~\kpc$ where the tidal streams would reside, the dark-matter only {\it ErisDark} simulation produces a
prolate halo with a minor/major axis ratio $c/a=0.45$ and an intermediate-major axis ratio $b/a=0.55$. In contrast, baryonic dissipation greatly increases the sphericity of the dark matter halo formed 
in the hydrodynamical {\it Eris} simulation, resulting in an oblate halo that is nearly axisymmetric in the $x-y$ plane with $b/a=0.98$. 
Perpendicular to the disk plane, the minor-major axis ratio in the {\it Eris} simulation increased to $c/a=0.65$. 
These results are 
consistent with those reported by \citet[][]{pillepich2014a}, who defined the halo axes following the iterative method described in \citet[][]{kuhlen2007a}. 
In both {\it Eris} and {\it ErisDark}, the gravitational potential is much rounder than the density, as expected from the Poisson equation, with a dark matter potential minor-to-major axis ratio of 
$\qzh \approx0.85$ in {\it Eris} and $\qzh \approx0.67$ in {\it ErisDark}. 

We highlight several features of Figure \ref{fig:shape}. The axis ratio of the dark matter halo in {\it Eris} varies with radius. Within $r\lesssim60~\kpc$ of the center, the halo grows rounder toward the galactic center. Baryonic infall redistributes the dark matter halo mass distribution. After including the contribution of baryons the total gravitational potential of {\it Eris}, the potential axis ratios increase at radii $r\gtrsim10~\kpc$. These radii lie well beyond the characteristic radii of the baryonic distribution in {\it Eris}, such that the baryons provide a point source-like contribution to the potential. In the innermost regions of the galaxy, the baryonic disk substantially flattens the total potential and lowers the sphericity of the total potential below that of the dark matter halo potential. The total potential flattening 
is $\qz\approx0.88$ in the {\it Eris} simulation at radii $r\gtrsim10~\kpc$.

\section{DYNAMICAL MODELING OF THE TIDAL STREAMS}
\label{sec:tidal streams}

Armed with an accurate representation of the gravitational potential in the {\it Eris} simulation, we can
model the evolution of tidal streams stripped from globular clusters orbiting in the simulated halo. We then
can compare the properties of these model 
streams with the observed Pal 5 and GD-1 streams, and check
for statistical consistency while marginalizing over the uncertainty in the globular cluster positions and velocities in the Milky Way halo.

\subsection{Streakline Method}
\label{subsec:streakline}
To model the tidal streams of Pal 5 and GD-1, we use the streakline method introduced by \citet{kupper2012a}. This method was used by \citet{kupper2015a} and \citet{bowden2015a} to constrain the potential of the Milky Way by modeling Pal 5 and GD-1, respectively. 
As the satellite moves in the potential of the host galaxy, test particles are released near the Lagrange points and their orbits integrated in the combined potential of the galaxy (computed from the SCF method) and satellite (approximated as a point mass). We ignore the mutual gravitational interaction of test particles and the gravitational force on the satellite from the particles. The streakline method proves significantly more computationally efficient than a full N-body simulation, and since the orbits of the stripped stars are 
assumed to be dynamically 
independent the computation is easy to parallelize. 

In our model,
the stripped stars separate from the progenitor globular clusters near the Lagrange points. We can define the Lagrange radius \citep{king1962a} as
\begin{equation}
r_t = \left(\frac{GM_{sat}}{\Omega^2-\partial^2\Phi/\partial r^2}\right)^{1/3},
\end{equation}
where $\Omega$ is the angular speed of the satellite in the galaxy potential, $\Phi$ is the gravitational potential of the host galaxy, and $M_{sat}$ is the mass of the satellite when the stars are released. 
For Pal 5 we strip stars from the inner Lagrange point to form a leading
stream, and from the outer Lagrange point to form a trailing stream. The tracers in the Pal 5 stream models are released every $\Delta t =8$ Myr, comparable with the $\Delta t=10$ Myr used by \citet{kupper2015a}.
For GD-1, following \citet{bowden2015a}
we release the particles at a distance $r=1.2r_t$ from the satellite
to prevent a large number of particles from being recaptured by the progenitor
cluster. The tracers in the GD-1 stream models are released every $\Delta t=2$ Myr. By the present day, more than 1000 tracer particles are left in the stream, comparable with the 500-1000 tracer particles in the
GD-1 stream simulated by 
\citet{bowden2015a}. 
Since no progenitor of GD-1 has been observed, we follow \citet{bovy2016a} in assuming the observed stream is leading and only
release stars around the inner Lagrange point when
modeling GD-1.

The initial velocities of the released stars are determined by
\begin{equation}
\label{eqn:v_initial}
\vec{v} = \vec{v}_0 + \delta \vec{v}
\end{equation}
where $\vec{v}_0$ reflects the radial and angular velocity of the satellite 
\citep{kupper2012a}, and $\delta \vec{v}$ represents a random 
component contributed by the velocity dispersion of the satellite. 
We assume that the internal velocity distribution of the satellite stars follows
a Gaussian distribution with zero mean and standard deviation $\sigma_v$. 
The mass of the satellite declines as the stars are gradually stripped, as represented by a constant mass loss rate $\dot{M}$. 
Since we only fit the mean orbit of the streams and do not attempt to model the density of stars along the stream, 
we release the stripped stars at a constant
rate. We do not attempt to model variations in the mass-loss rate from the globulars that might subtly affect their orbits. 
Test particles do not 
correspond to individual stars, and the mass loss rate of the satellite
does not scale with the rate at which we release particles.
In total, our model has eight free parameters for the progenitor cluster including its average mass loss rate,  present-day mass, and
 six current phase space coordinates.
For Pal 5, constraints on the present-day progenitor cluster phase space coordinates and the relation between mass and mass loss rate
allow us to reduce the number of free parameters 
to four including the distance, proper motions, and mass of the cluster.

Starting with the location and velocity of the progenitors at the present, we integrate each progenitor backward by $\Delta t = 4$ Gyr
to set the initial conditions.
This chosen timescale is long enough for 
stripped stars to cover the whole Pal 5 and GD-1 streams by the present day, and cautiously exceeds that used by  previous analyses \citep[e.g.,][]{bovy2016a}. Increasing this timescale would increase the length of the simulated
stream, but will have little influence on the stream 
properties within the extent of the data.
We then integrate orbits forward in time using a leapfrog integrator \citep[e.g.,][]{press1986a}. 
We treat any stars that fall within half of the Lagrange radius of their parent globular cluster as "re-captured'', and remove them from the simulation. 
Once the systems evolve to the present day, we use local regression with a tri-cubic weighting function to calculate the mean track of the stellar
orbits along the stream.

Owing to the non-zero velocity dispersion for the stars released from the globular clusters, some particles will move away from the main stream orbit and introduce
noise into its determination.
Since both the Pal 5 and GD-1 streams are dynamically cold, 
we reduce the particle noise by reducing the random velocity component in Equation
\ref{eqn:v_initial}. 
We fix $\delta \vec{v} = 0$ for Pal 5 (following \citealt{kupper2015a}) 
and set $\sigma_{v} = 0.25~\kms$ for GD-1.  
We have verified that reducing the
random velocity component does not substantially
affect the results
of our analysis.

\subsection{ Milky Way-like Potential Models}

\begin{table}[tp]
    \renewcommand\arraystretch{1.5}
    \centering
    \caption {\label{table:BovyModel} Parameters of the spherical halo model}
    \begin{tabular}{ p{1.5cm}<{\centering} p{3.2cm}<{\centering} p{2.7cm}<{\centering} }
    \hline
    Component & Parameter & Value\\
    \hline
    Halo & scale length $r_s$ & $17.4~\kpc$\\
     & mass $M_{\mathrm{DM}}(<20~\kpc)$ & $1.1\times10^{11}~\Msun$\\
    Bulge & total mass $M_b$ & $0.3\times10^{10}~\Msun$\\
    Disk & total mass $M_d$ & $7.0\times10^{10}~\Msun$\\
     & scale length $R_d$ & $3.0~\kpc$\\
     & scale height $Z_d$ & $0.28~\kpc$\\
    \hline
    \end{tabular}
\end{table}

We compare three different models for the gravitational potential of Milky Way-like galaxies in our stellar stream simulations.
In the first, we use the SCF expansion of the {\it Eris} potential described in Section \ref{subsec:model}. We assume the potential to be
azimuthally symmetric and mirror symmetric about the disk plane, leaving only 66 terms in our SCF expansion.
In our second model, we use an SCF expansion of the {\it ErisDark} potential. Although the dark matter halo
in this simulation is prolate and triaxial, we take an azimuthal average 
resulting in an oblate axisymmetric 
spheroid with a potential flattening of $\qz\approx0.8$.
We make this assumption to match constraints from studies that suggest the axis ratio of the halo in the disk plane is close to unity
\citep{bovy2014b, bovy2015a}. As we demonstrate below, the {\it ErisDark} simulation does not 
reproduce well the stellar stream data and 
using a prolate model would just further exacerbate this failure.
In our third model, we assume a 
potential similar to the best fit found by \citet{bovy2016a} including
a Miyamoto-Nagai disk and a spherical bulge. However, instead of adopting a triaxial NFW dark matter halo with $c/a=1.05$ we
use a spherical NFW halo with the same scale length and normalization. 
The spherical halo greatly simplifies the calculation
and does not have a strong effect on the resulting stream orbits relative to a nearly spherical triaxial ellipsoid. The detailed parameters of this case can be found in Table \ref{table:BovyModel}.
In the following analysis we treat results from the spherical halo model as a benchmark for evaluating the success of {\it Eris}, and compare {\it Eris} and {\it ErisDark} to study the importance of baryonic physics for
altering the potential in which we evolve the stream orbits.

\subsection{Data}
To evaluate the {\it Eris} potential shape as a 
template 
for a Milky Way-like galaxy, we perform an analysis
similar to that presented by \citet{bovy2016a}. We utilize constraints from both the Pal 5
and GD-1 stellar streams. For both datasets, to transform between the Celestial coordinates and Galactocentric coordinates we fix the solar galactocentric radius  $R_0=8.0~\kpc$, the solar vertical position
in the Galactic disk $z_0=25~\pc$, and the solar velocity with respect to the local standard of rest 
 $[V_R,\ V_T,\ V_Z]=[-11.1,\ 12.24,\ 7.25]~\kms$ \citep{schonrich2010a}. 

\subsubsection{Pal 5}

\begin{figure*}[htp]
\centering
\includegraphics[width=\linewidth, clip]{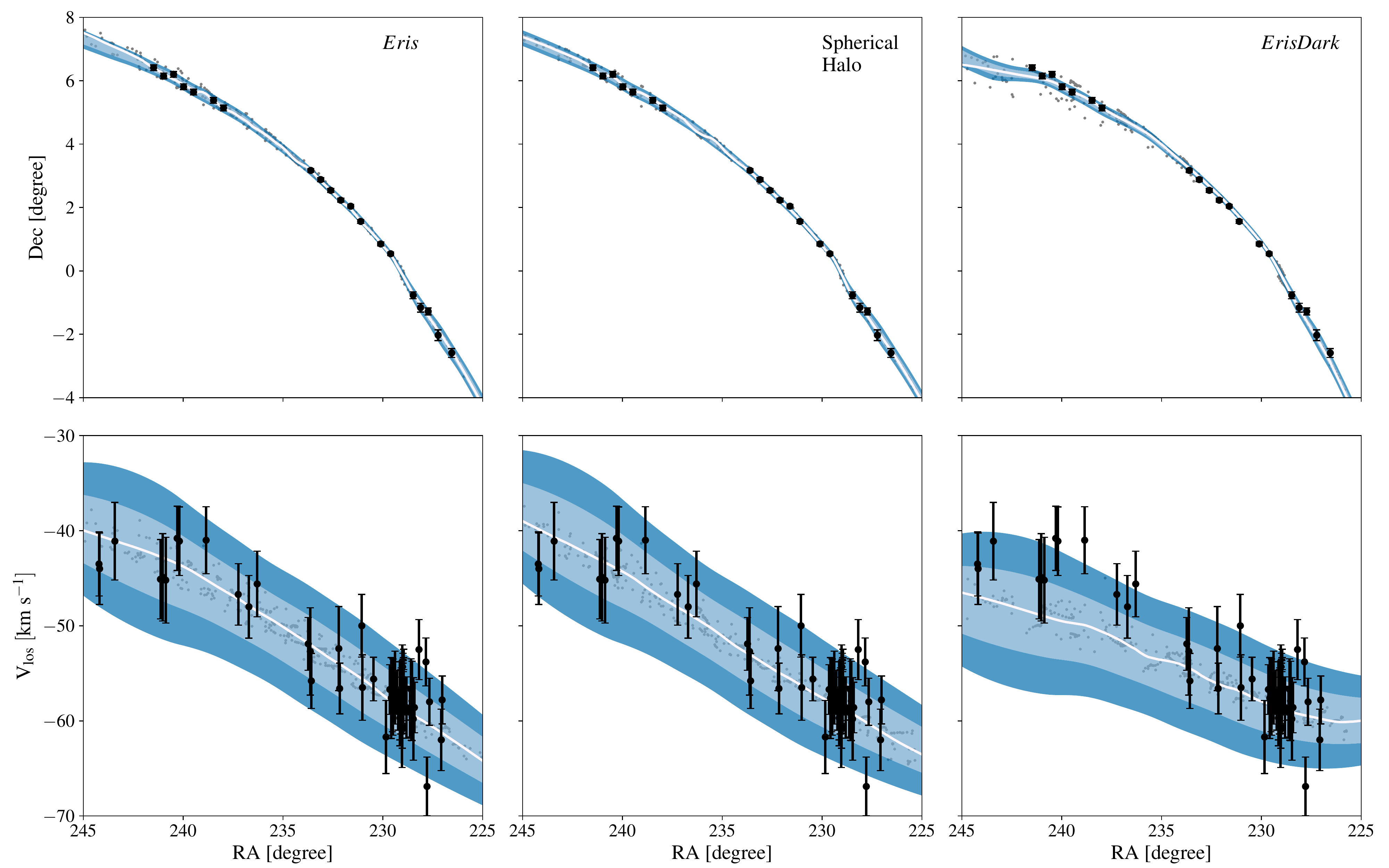}
\caption{Tidal stream data (points with error bars) for the disrupting Pal 5 globular cluster, along with the {\it Eris} halo (left panels), best-fit spherical halo (middle panels), and the {\it ErisDark} halo (right panels) models. The models attempt to reproduce
the path of the tidal stream on the sky by matching the right ascension $\alpha$ and declination $\delta$ of the
stars (upper panels) and how their line-of-sight velocities $V_{los}$ depend on right ascension (bottom panels).
Our Bayesian analysis uses a Monte Carlo approach, varying the line of sight position, proper
motions, and mass of Pal 5 as inputs into a streakline model for generating the tidal streams. Tens of thousands of
potential stream orbits are considered and compared with the data shown in this figure. The bands show the
inner 67\% and 95\% credibility intervals on stream orbits in the halo potentials.
The solid lines present the maximum likelihood stream track in each potential, while the small gray points show 
 a representative subset of the present day test star particles associated with the maximum likelihood parameters.
Both {\it Eris} and the spherical halo allow for orbits that are highly
consistent with the Pal 5 data, even though the total potential flattening is quite different for these two models: $\qz=0.88$ for {\it Eris} and $\qz=0.97$ for spherical halo model at the location of Pal 5 cluster.
The dark matter-only simulation {\it ErisDark} does
not produce a comparably good model of the Pal 5 data.}
\label{fig:pal5}
\bigskip
\end{figure*}

Figure \ref{fig:pal5} shows the Pal 5 tidal stream data. 
We use the position of the stream stars on the sky in right ascension (RA) $\alpha$ and 
declination $\delta$, the 
six phase-coordinate location of the Pal 5 cluster from \citet{fritz2015a}, and the line of sight velocity measurement of the Pal 5 cluster from \citet{kuzma2015a} to constrain the model stream orbits. We treat the line-of-sight velocity of each star as equal to the mean value of the tidal stream at the RA of the star, following \citet{bovy2016a}.
The relation between the mass $M_{cl}$ and the mass loss rate $\dot{M}$ for Pal 5, $\dot{M}/M_{cl}=(-0.7\pm0.2)~\mathrm{Gyr}^{-1}$, is taken from \citet{odenkirchen2003a}. We fit both the leading arm and trailing arm of Pal 5 simultaneously.

\begin{figure*}[htp]
\centering
\includegraphics[width=\linewidth, clip]{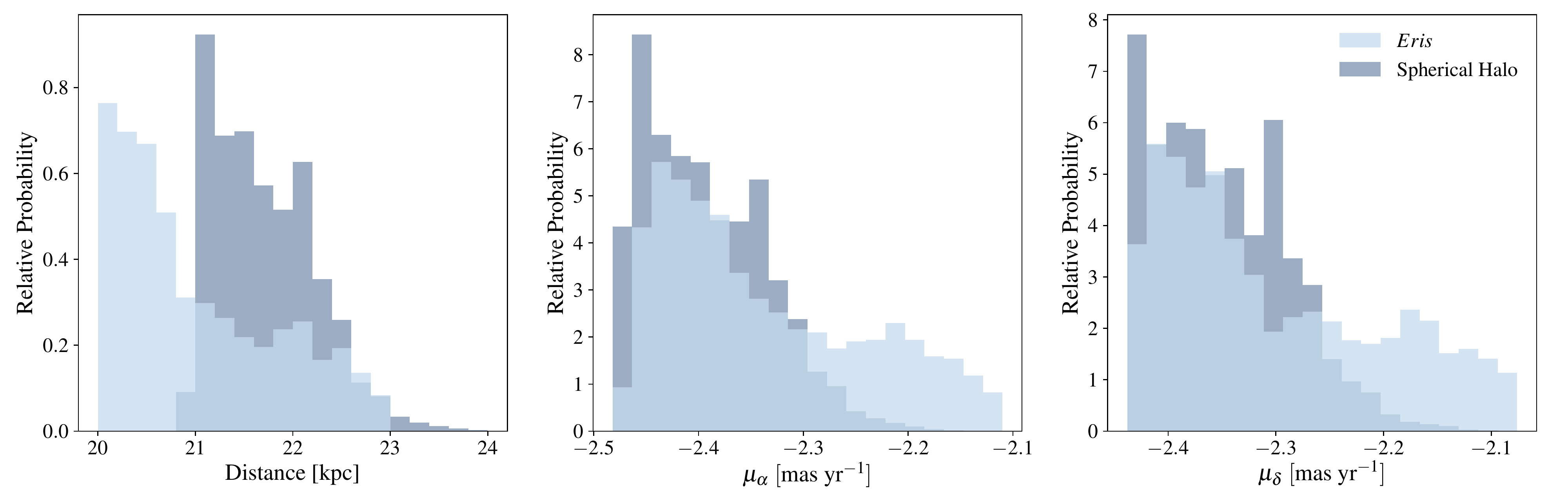}
\caption{Marginalized 1-D posterior probability distributions of the three Pal 5 cluster parameters. The priors on these parameters are set using observational constraints (see Section \ref{subsec:bayesian} and Table \ref{table:priorP}). The light blue color displays the normalized posterior distribution for the {\it Eris} potential, while the dark blue color shows the spherical halo model results.}
\label{fig:Pal5posterior}
\bigskip
\end{figure*}

\subsubsection{GD-1}

\begin{figure*}[htp]
\centering
\includegraphics[width=\linewidth, clip]{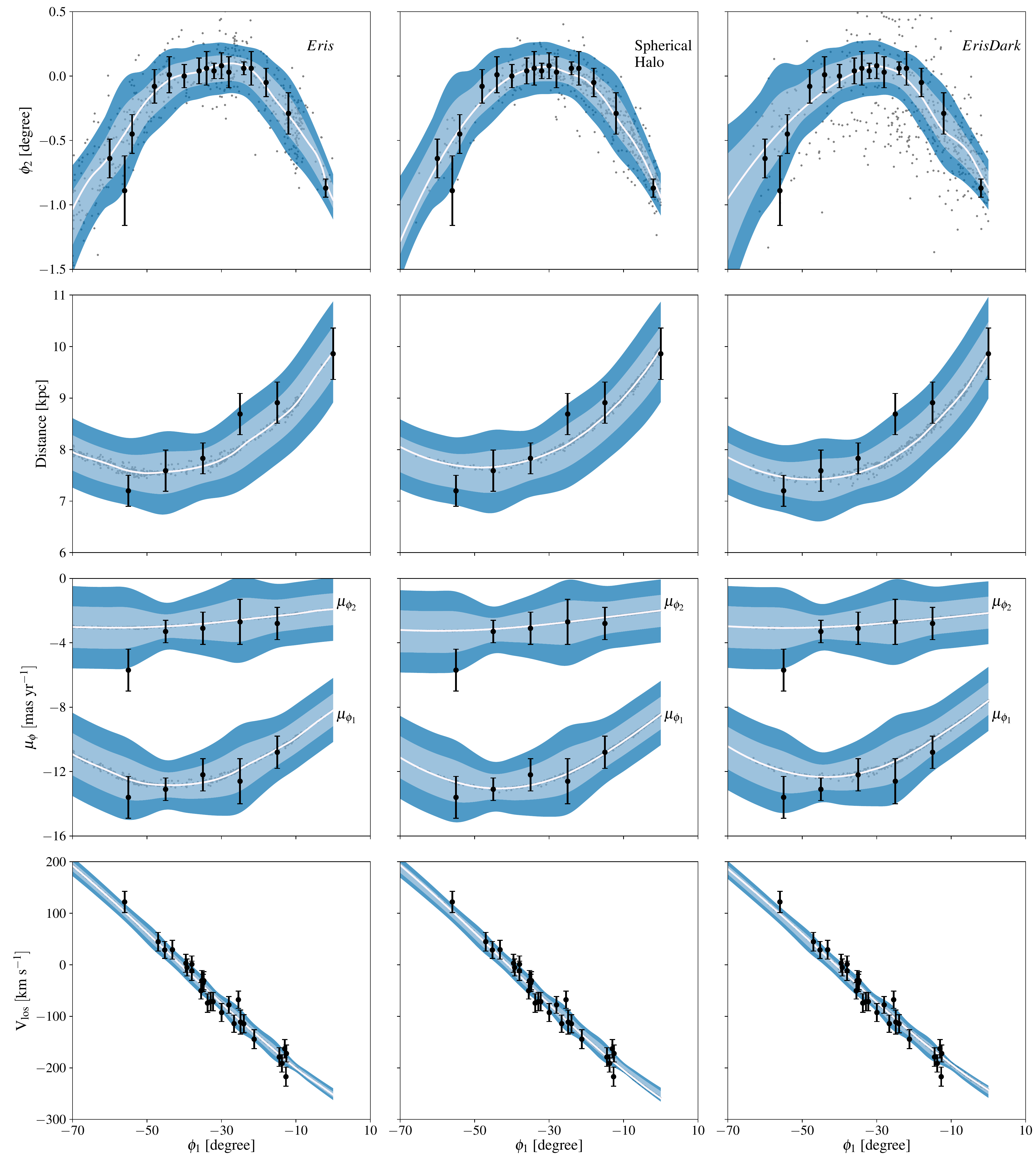}
\caption{Tidal stream data (points with error bars) for the GD-1 stream, along with the {\it Eris} halo (left panels), best-fit spherical halo (middle panels) models, and {\it ErisDark} halo (right panels). 
Symbols, bands, and lines are defined as in the
Figure \ref{fig:pal5} caption.  A representative
subset of the stellar particles are shown (small gray points).
The models attempt to reproduce
the path of the tidal stream on the sky by matching the star positions $\phi_1$ and $\phi_2$
along the stream (upper most panels), the line-of-sight-distance along the stream (second panels from top), the
stellar proper motions along the stream (third panels from top),
and how their line-of-sight velocities $V_{los}$ depend on position along the stream (bottom panels).
Both {\it Eris} and the spherical halo allow for orbits that are highly
consistent with the GD-1 data, but {\it ErisDark} does
not.}
\label{fig:gd1}
\bigskip
\end{figure*}

Figure \ref{fig:gd1} shows the GD-1 tidal stream data as points with error bars.
The six-dimensional phase-space data from \citet{koposov2010a} comprise our constraints, and consist
of the positions $[\phi_1, \phi_2]$ corresponding to a coordinate system oriented along and perpendicular to
the stellar stream, the line-of-sight distance along the stream, the proper motions along and perpendicular to the stream, and the line-of-sight velocity of each star along the stream. Similar to Pal 5, the line-of-sight velocity measurement of each star is treated as the mean value of the stream as a function of position along the
sky.

\subsection{Bayesian Analysis}
\label{subsec:bayesian}

Given the
{\it Eris} potential, which we treat as fixed according to our SCF model, and
a set of parameters $\vec{\theta}$ (some of the six phase-space coordinates of the progenitor),
we can calculate the mean orbit of the stream following the streakline method described in Section \ref{subsec:streakline}. We then use a Bayesian analysis to determine the likelihood of the stellar stream
data given the model, and locate the maximum likelihood of the progenitor properties that produce
streams most consistent with the available data.

We define our log-likelihood as:
\begin{equation}
    \log\mathcal{L(\vec{\theta})}=-\sum_{i,j}\frac{\left[x_{data,ij}-f_{ij}(\vec{\theta})\right]^2}{2\sigma_{ij}^2},
\end{equation}
\noindent
where $i$ denotes different data types (for example, the position or line-of-sight velocity), 
$j$ runs over the number of data points of the same type, 
$f_{ij}(\vec{\theta})$ represents the model prediction for the data given parameters $\vec{\theta}$, $x_{data,ij}$ are the data points, 
and $\sigma_{ij}$ are the corresponding uncertainties for each observed star. Since the line-of-sight velocity data are measurements of individual stars rather than the whole stream, we must account for the
velocity dispersion of the stars about the stream. 
We therefore add a stream velocity dispersion $\sigma_{sys}$ to 
the uncertainty in each star's velocity in quadrature:
\begin{equation}
    \sigma_{ij} = \sqrt{\sigma_{obs, ij}^2+\sigma_{sys}^2}
\end{equation}
\noindent
We determine the stream velocity dispersion by fitting a cubic polynomial to the stellar data velocities as a proxy for
the stream velocity and then calculating their variance about this fit. We find stream velocities dispersions of 
$\sigma_{Pal5,sys} = 2.26~\kms$ and $\sigma_{GD1,sys} = 15.08~\kms$.
To provide the sampling for our analysis, calculate the Bayesian evidence, and
compute the posterior distribution of each model, we use {\it Multinest}
\citep{feroz2009a}.

\begin{table}[tp]
    \renewcommand\arraystretch{1.5}
    \caption{\label{table:priorP}Model Priors for Pal 5}
    \centering
    \begin{tabular}{ p{2cm}<{\centering} p{2.7cm}<{\centering} p{2.7cm}<{\centering} }
    \hline
     & Minimum & Maximum \\
    \hline
    Distance & $20~\kpc$ & $24~\kpc$\\
    $\mu_{\alpha}$ & $-2.482~\masyr$ & $-2.110~\masyr$\\
    $\mu_{\delta}$ & $-2.438~\masyr$ & $-2.076~\masyr$\\
    Mass & $5\times10^3~~\Msun$ & $2.5\times10^4~~\Msun$\\
    \hline
    \end{tabular}
\end{table}

\begin{table}[tp]
    \renewcommand\arraystretch{1.5}
    \caption{\label{table:priorG}Model Priors for GD-1}
    \centering
    \begin{tabular}{ p{2cm}<{\centering} p{2.7cm}<{\centering} p{2.7cm}<{\centering} }
    \hline
     & Minimum & Maximum \\
    \hline
    $\phi_1$ & $0^\circ$ & $20^\circ$ \\
    $\phi_2$ & $-3.5^\circ$ & $0^\circ$ \\
    Distance & $9.5~\kpc$ & $15.5~\kpc$ \\
    $\mu_{\phi_1}$ & $-9~\masyr$ & $-4~\masyr$\\
    $\mu_{\phi_2}$ & $-2.2~\masyr$ & $-1.2~\masyr$\\
    $V_{los}$ & $-380~\kms$ & $-250~\kms$ \\
    Mass & $10^4\Msun$ & $10^5\Msun$\\
    Mass-loss Rate & $0~\Msun~\Myr^{-1}$ &  $20\ ~\Msun~\Myr^{-1}$\\
    \hline
    \end{tabular}
\end{table}

Following \citet{bovy2016a}, we fix the 
celestial coordinates of Pal 5 at 
$\alpha=229.018^\circ$, $\delta=-0.124^\circ$, and fix
the line-of-sight velocity to $V_{los}=-58.7~\kms$.
We assume a flat prior on the proper motion within the observed uncertainty, 
as shown in Table \ref{table:priorP}.
The measurements of the distance of the Pal 5 cluster range over $20.9-23.2~\kpc$ \citep{harris1996a, vivas2006a, dotter2011a}, and we therefore assume a flat prior for the line-of-sight distance of $D\in[20~\kpc,24~\kpc]$.
Mass estimates for Pal 5 vary, ranging from $5\times10^3~\Msun$ to $2.2\times10^4~\Msun$ \citep{odenkirchen2002a, kupper2015a}, 
and we adopt a flat prior of $M_{cl}\in[5\times10^3~\Msun,2.5\times10^4~\Msun]$.
For GD-1, since no progenitor has been identified we assume a flat prior on all six progenitor phase space coordinates, the mass, and the mass loss rate. We first performed a test run with a broad prior, and then reset the prior based on the marginalized one-dimensional posterior distribution of the test run.
We provide a complete listing of the model priors for Pal 5 and GD-1 in Tables \ref{table:priorP} and \ref{table:priorG}, respectively.

\section{RESULTS}
\label{sec:results}

\begin{table}[tp]
    \renewcommand\arraystretch{1.5}
    \caption{\label{table:evidence}log-Evidence for Milky Way-like Potential Models}
    \centering
    \begin{tabular}{ p{2.5cm}<{\centering} p{2.5cm}<{\centering} p{2.5cm}<{\centering} }
    \hline
    & Pal 5 & GD-1\\
    \hline
    Spherical Halo & $-43.30 \pm 0.09$ & $-35.08 \pm 0.11$\\
    {\it Eris} & $-42.20 \pm 0.09$ & $-36.10 \pm 0.11$\\
    {\it ErisDark} & $-81.39 \pm 0.11$ & $-49.90 \pm 0.14$\\
    \hline
    \end{tabular}
\end{table}

\begin{table}[tp]
    \renewcommand\arraystretch{1.5}
    \caption{\label{table:chi2}Best $\chi^2$ for Milky Way-like Potential Models}
    \centering
    \begin{tabular}{ p{2.5cm}<{\centering} p{2.5cm}<{\centering} p{2.5cm}<{\centering} }
    \hline
    & Pal 5 & GD-1\\
    \hline
    Spherical Halo & $31.57$ & $19.01$\\
    {\it Eris} & $31.47$ & $19.11$\\
    {\it ErisDark} & $65.77$ & $25.17$\\
    \hline
    \end{tabular}
\end{table}
In this Section we present the results of the dynamical modeling and Bayesian analysis of the tidal streams Pal 5 and GD-1 in our models for the Milky Way potential.

\subsection{Pal 5}

Figure \ref{fig:pal5} shows the 68\% and 95\% credibility intervals for the mean tidal stream orbits computed in the
{\it Eris} (left panels), spherical halo (middle panels), and {\it ErisDark} (right panels) models for the declination $\delta$ (upper panels) and line-of-sight velocity $V_{los}$ (bottom panels) as a function of right ascension $\alpha$. 
To compute the credibility intervals at each position along the stream, likelihood samples from the Bayesan analysis were used to compute the
posterior predictive distributions for declination
$\delta$ or line-of-sight velocity $V_{los}$ as
a function of right ascension $\alpha$. In calculating 
the posterior predictive distribution, we assigned uncertainties to regions between the data points 
by interpolation.
As Figure \ref{fig:pal5} demonstrates, both the {\it Eris} and spherical halo models allow for stellar 
streams that appear highly consistent with the available data. We find this compatibility remarkable given
that the spherical model adopts the best-fit model of \citet{bovy2016a}, with a halo mass and
shape selected from a full Bayesian analysis to provide maximal consistency with these data,
whereas the {\it Eris} potential was in no way selected to reproduce the Pal 5 stream.

To quantify the relative agreement, we present the log-evidence of 
both the {\it Eris} and spherical halo models for the Pal 5 stream
in Table \ref{table:evidence} and the $\chi^2$ of the best fit parameters in Table \ref{table:chi2}. The Bayesian evidence is defined as the average of the likelihood over 
the prior, and can be written as
\begin{equation}
\mathcal{Z}=\int{\mathcal{L}(\theta)\pi(\theta) d\theta}
\end{equation}
where $\theta$ are the parameters in the model, $\mathcal{L}(\theta)$ is the likelihood, and $\pi(\theta)$ is the prior.
{\it Eris} provides the best model, but the spherical halo model allows for streams that show
comparable consistency with the data. Given that the spherical halo model parameters remained fixed
during our analysis this comparison is formally correct, 
but we note that the
additional flexibility of allowing the halo mass and radius to vary could produce
a somewhat better log-evidence than shown in Table \ref{table:evidence}. Nonetheless,
Figure \ref{fig:pal5} and Table \ref{table:evidence}
suggest that the {\it Eris} and spherical halo
models work comparably well. In Table \ref{table:evidence}, we
also present the log-evidence for models generated from the dark matter-only {\it ErisDark}
simulation.
{\it ErisDark} performs poorly, which is unsurprising given the strongly aspherical potential
without dissipative effects or a large central baryonic mass concentration. 
The end of the leading arm and the trailing arm of the best fit {\it ErisDark} stream both deviate from the observed data.
Stars that lie at the end of the trailing are typically the first
stripped and travel the farthest away from the cluster, and
therefore are quite sensitive to the shape of the host galaxy potential.

Figure \ref{fig:Pal5posterior} presents the one dimensional marginalized posterior distribution of the Pal 5 distance and proper motions in the {\it Eris} and 
the spherical halo models. As expected, these parameters have quite different posteriors in the
{\it Eris} and spherical halo models, reflecting
differences in their potential shapes. The overlap
between the two posterior distributions is comparable to
the width of the spherical halo posterior with $\sigma_{\mu}\approx 100 \muasyr$, suggesting that future data more precise than this level would be required to distinguish between
the two models.
We note that owing to the strong correlation between distance and proper motions the
posterior distribution for the distance to Pal 5 cluster in the spherical halo model truncates at $D \approx 21 \kpc$, corresponding to a proper motion near the
lower limit of the proper motion prior.
However, we have checked that increasing
the proper motion 
prior range does not substantially change these
results.

\subsection{GD-1}

Figure \ref{fig:gd1} shows the the 68\% and 95\% credibility intervals for tidal stream orbits computed in the
{\it Eris} (left panels), spherical halo (middle panels), and {\it ErisDark} (right panels) models models. The data for
the GD-1 stream are presented in the rotated spherical coordinates ($\phi_1$, $\phi_2$) oriented parallel and perpendicular to the direction of the stream orbit. Shown are
the positions $\phi_2$, line-of-sight distance, proper motions, and line-of-sight velocities
as a function of the $\phi_1$ coordinate for each model. Only a representative subsample of 500 tracers are plotted, while more than 1000 tracers are used to calculate the stream centerline. As with Pal 5, we show the
marginalized 68\% and 95\% credibility intervals for each orbital parameter as bands, and
the maximum likelihood orbit as a solid line. Both  the {\it Eris} and spherical halo potential models provide comparable
fits given their complexity. In contrast, the best fit stream in
the {\it ErisDark} potential encounters difficulty at large $\phi_1$, and
produces broader streams than the other potential models. 
Table \ref{table:evidence} and Table \ref{table:chi2} provide
the log-evidence and best $\chi^2$ for each model, showing that the spherical
halo is sightly favored
to {\it Eris}, but that the difference in log-evidence is less than for Pal 5. The
{\it ErisDark} model performs substantially poorer. 

\section{DISCUSSION}
\label{sec:discussion}

Our analysis demonstrates that the gravitational potential of the
simulated
{\it Eris} galaxy permits the formation of tidal streams that
closely resemble Pal 5 and GD-1.
We emphasize that we did not select {\it Eris} among a range of models,
nor did we tune its physical prescriptions for the purpose
of producing a simulated halo that would allow for realistic stellar streams.
{\it Eris} simply provides a close Milky Way analog that serves as
a model gravitational potential for comparing with observed Milky Way
stellar streams.
We did not attempt to find a statistically optimal model potential in the
maximum likelihood sense, as the {\it Eris} potential remained fixed
throughout our tests. As we show below, with the greater flexibility of allowing
the halo shape or disk mass to vary we can find a somewhat closer match to either the Pal 5 or GD-1 stellar stream data than {\it Eris} provides. However,
{\it Eris}, a simulated galaxy formed
fully self-consistently in a complete cosmological context, works very
well as a Milky Way-like potential model as probed by the Pal 5 and GD-1
streams. Importantly, this result suggests that structure formation in
the $\Lambda$CDM cosmology can produce galaxies whose halos closely resemble
that of the Milky Way. The $\Lambda$CDM structure formation process can accomplish this success without resorting to additional physics beyond those associated with standard CDM and the dissipational processes associated with baryons.

\begin{figure*}[t] 
\centering
\includegraphics[width=\linewidth, clip]{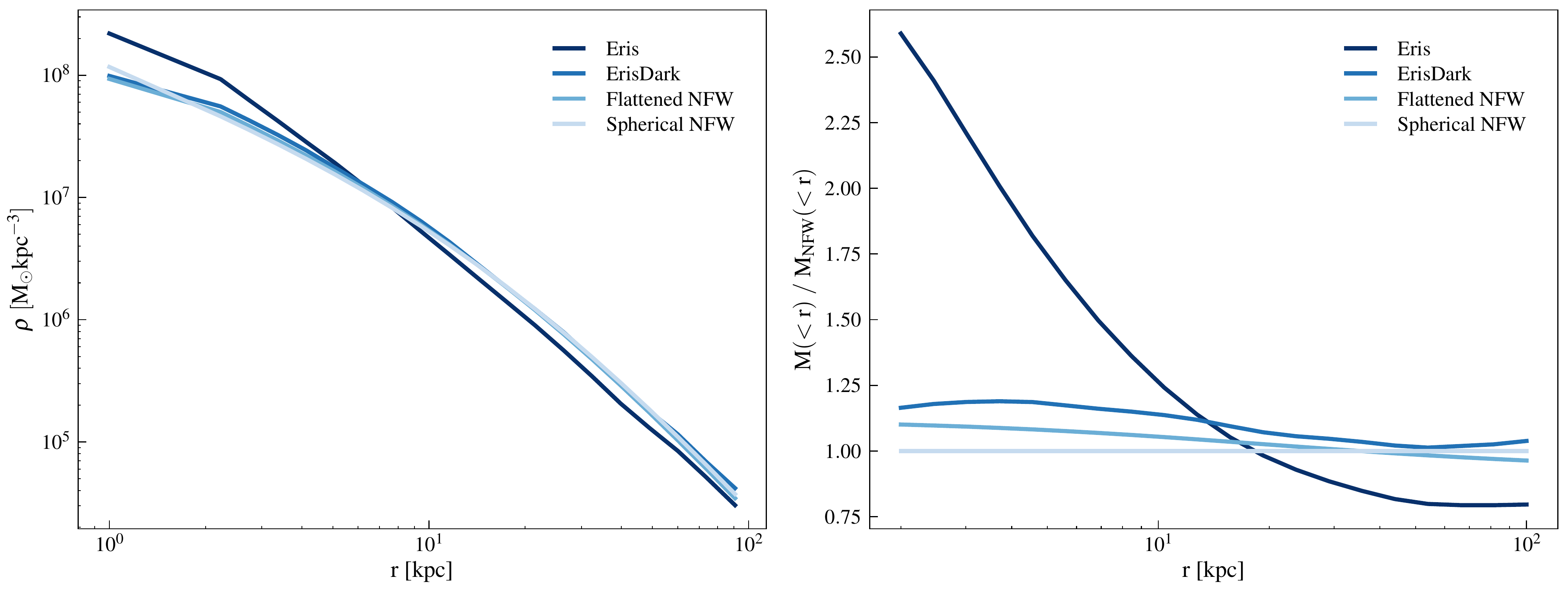}
\caption{Comparison of the average radial 
density profile ($\rho(r)$; left) and interior mass distribution ($M(<r)$; right) of dark matter halo models used in
computing stream orbits. In the right panel, the interior mass distribution of each model is normalized by the NFW interior mass profile. Relative to a spherical NFW halo (lightest blue), baryonic dissipation increases the central concentration of the {\it Eris} halo mass distribution (darkest blue). As expected, {\it ErisDark} (dark blue) and a flattened NFW halo (light blue) with $\qzh=0.85$ (chosen to have oblateness similar to {\it Eris}) are slightly more centrally-concentrated relative to a spherical NFW.}
\label{fig:mass_profile}
\bigskip
\end{figure*}

\subsection{Effects of Baryonic Dissipation}
\label{subsec:dissipation}

The comparison of the dark matter halos in the {\it Eris} and {\it ErisDark} simulations
in Section \ref{sec:shape} demonstrated that baryonic dissipation greatly increased
the sphericity of the halo. However, the {\it Eris} halo is still substantially oblate.
The comparable performance of the oblate {\it Eris} halo and a spherical NFW halo suggests
that differences between the potential form in the two halo models, in addition to the halo flatness, influences the stream
orbits. We now examine how baryonic dissipation changes
the mass distribution in {\it Eris}.

Figure \ref{fig:mass_profile} shows the average radial density profile $\rho(r)$ and interior
mass distribution $M(<r)$ for {\it Eris}, {\it ErisDark}, and a spherical NFW. The prolate halo
in the dissipationless {\it ErisDark} simulation is slightly more centrally concentrated than
a spherical NFW halo. The much poorer performance of {\it ErisDark} in reproducing the streams
relative to the spherical NFW halo therefore
results primarily from the flattened {\it ErisDark} halo shape, rather than large
differences in their radial mass distribution. 
In contrast, baryonic
dissipation redistributes the mass of {\it Eris} inward compared with a spherical NFW halo and
rounds the halo relative to {\it ErisDark}. 

Since {\it Eris} performs comparably to a 
spherical NFW halo but has a flatness intermediate between {\it ErisDark} and spherical,
we performed additional tests to explore the relative roles of radial mass re-distribution or rounding of the
halo shape in improving the {\it Eris} performance.
We revisited our previous analysis using a spheroidal NFW halo with a 
flattening $\qzh=0.85$. This model mimics the {\it Eris} halo shape, but uses the
more extended NFW halo mass distribution. In terms of the mass $M_s$ within the scale 
radius $r_s$, the potential of a spheroidal NFW halo can be written as
\begin{equation}
\label{equ:potentialq}
\Phi^h(m)=-\frac{GM_s}{m}\log(1+m/r_s) 
\end{equation}
\noindent
where
\begin{equation}
m^2=x^2+y^2+\frac{z^2}{{\qzh}^2}.
\end{equation}
\noindent
If $\qzh=1$, then Equation \ref{equ:potentialq} reduces
to the spherical NFW potential. The mass distribution of this
flattened NFW model is also plotted in Figure \ref{fig:mass_profile}, showing
as expected
that a spheroidal NFW model is slightly more centrally-concentrated than an
spherical NFW model.

Using this flattened NFW potential model with $\qzh=0.85$, we
perform the same Bayesian analysis described in Section \ref{subsec:bayesian}.
For the Pal 5 stream, we find a log-evidence of $\log \mathcal{Z}=-41.55\pm0.09$, 
slightly better than for {\it Eris} or a spherical NFW halo. For the GD-1 stream, 
we find a log-evidence of $\log \mathcal{Z}=-38.17\pm0.12$, 
worse than for {\it Eris} or a spherical NFW halo. The performance of the spheroidal NFW
halo when combining both the Pal 5 and GD-1 constraints is worse than for either {\it Eris} or a spherical NFW halo.
From these results we infer that the mass re-distribution by baryonic infall is somewhat more influential than the
rounding of the halo shape in improving the performance of {\it Eris}, but that
the relative importance of these two effects may vary with radius.

We also note that the degree of dissipation in {\it Eris} should be model-dependent
and in principle the effect of baryonic dissipation could be stronger with, e.g.,
 a different
feedback prescription or in a galaxy with a different formation time
than {\it Eris}. \citet{kazantzidis2010a} used
simulations to show that the amount of sphericalization not only depends
on the mass of the forming disk, but also on its alignment with halo.
The effect of dissipation in rounding the halo is strongest if the 
symmetry axis of the disk is aligned with the major axis of the dark halo. 
Since the axis of the disk in {\it Eris} is aligned with the minor axis of the halo, and the {\it Eris} disk mass is smaller than the Milky Way, 
it is plausible that the effect of baryonic dissipation on the halo sphericity
could be stronger in the real
Milky Way.

\subsection{Comparison to Previous Studies}
\label{subsec:comparison}

Several previous analyses studied the development of Milky Way
stellar streams, and used observed streams as constraints on the 
Milky Way halo shape. Our formalism enables us to compare straightforwardly
with previous analyses and, in some cases, attempt a reproduction of their
results. In all cases, we find broad consistency where we can compare. This
agreement helps to validate both the findings of previous studies and the
Bayesian methodology presented in this work.

\begin{figure}[t]
\centering
\includegraphics[width=3.4in, clip]{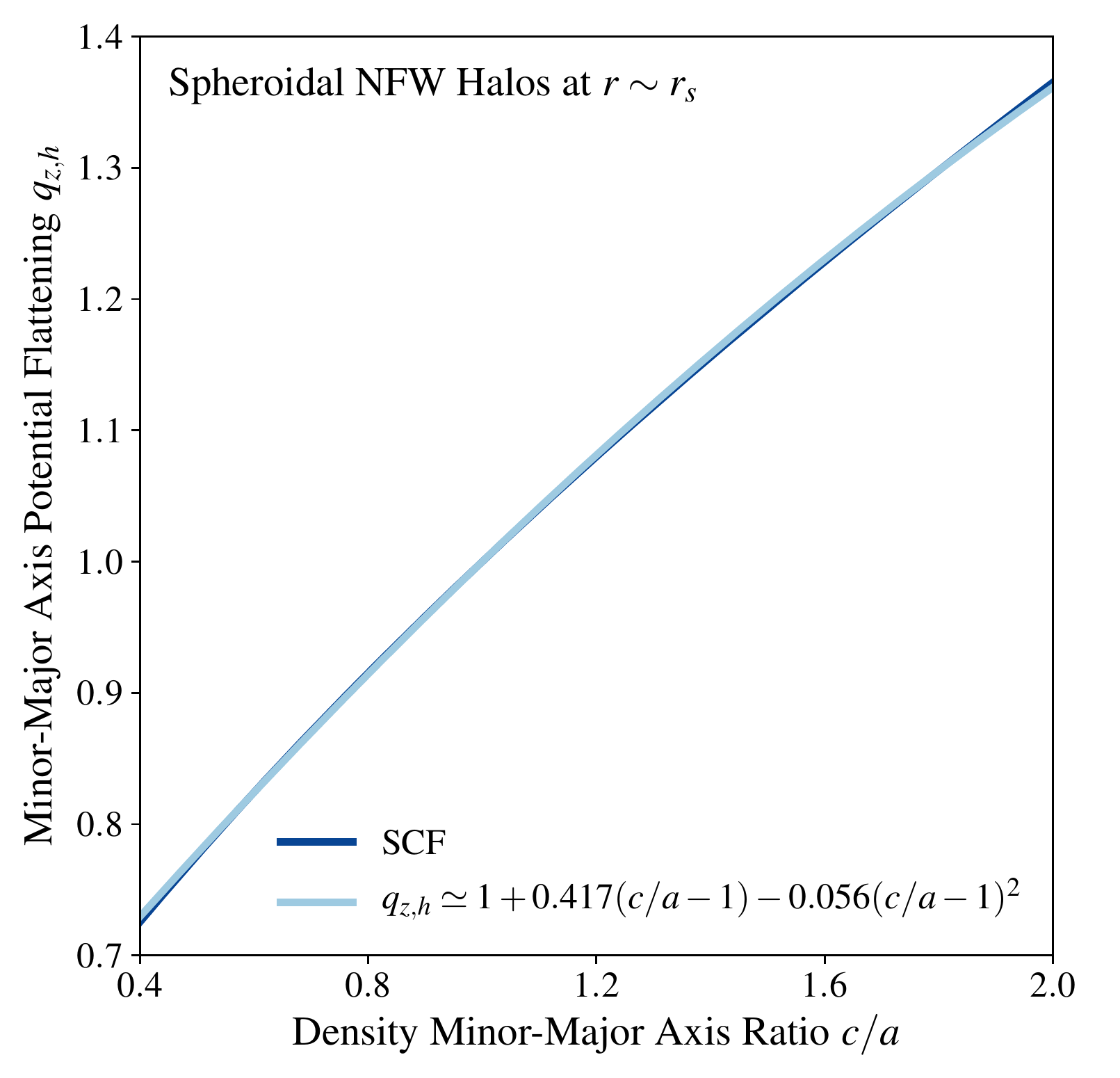}
\caption{Potential flattening versus axis ratio of the density field for a flattened NFW profile. The dark blue color shows our measured relation, in which $\qzh$ is measured by fitting ellipse to the potential field at the $r_s$ radius, and the potential field is calculated using our basis expansion.  The light blue color displays our best fit line in the form of Equation \ref{eqn:q_vs_ca}. This relation is only satisfied near the inner halo ($r \sim r_s$) of a NFW profile with a constant density axis ratio.}
\label{fig:q_vs_ca}
\bigskip
\end{figure}

\subsubsection{Halo Potential vs. Density Shape}

In what follows, we compare with studies that constrain either the halo
potential flattening $\qzh$ or the halo density flattening expressed
in terms of the minor-to-major axis ratio $c/a$. To make  these
comparisons clearer, 
we link the shape of the density field to potential flattening by calculating the potential field of a spheroidal NFW density profile. We model the potentials of flattened NFW profiles with different axis ratios with our basis expansion, and then fit ellipses to the isopotential contours to measure the potential flattening. 
Figure \ref{fig:q_vs_ca} shows this simple correspondence between
$\qzh$ and $c/a$. We find that a quadratic function can fit well this relation as
\begin{equation}
\label{eqn:q_vs_ca}
\qzh=1+\eta(c/a-1)-\xi(c/a-1)^2,
\end{equation}
\noindent
with the numerical constants $\eta\approx0.417$ and $\xi\approx0.056$.
Note that we measured the potential shape at the NFW profile scale radius $r = r_s$, and since the potential flattening is radius-dependent this relation remains only satisfied in the inner halo with $r \sim r_s$. As expected, a spherical density
profile results in a spherical potential. Note that Figure \ref{fig:q_vs_ca} 
and Equation \ref{eqn:q_vs_ca} are consistent with connection between $\qzh$
and $c/a$ in the {\it Eris} halo, as shown in Figure \ref{fig:shape}.
Equation \ref{eqn:q_vs_ca} also provides a rule of thumb for comparing the relative
statistical uncertainties $\sigma_q$ or $\sigma_{c/a}$ on $\qzh$ or $c/a$, and via the error propagation formula we have
\begin{equation}
\sigma_{c/a} \approx \frac{\sigma_{q}}{\sqrt{\eta^2 - 4\xi(\qzh-1)}}.
\end{equation}
\noindent
Near $\qzh\approx c/a \approx 1$, we have roughly
that $\sigma_{c/a}/\sigma_{q}\approx\eta^{-1}\approx2.4$ for NFW halos.

\subsubsection{Comparison to Previous Streakline Models}

The {\it Eris} simulation produced a Milky Way-like galaxy with an oblate dark matter halo and a total potential flattening $\qz=0.88$, and our results suggest that the {\it Eris} potential can produce tidal streams remarkably consistent with the Pal 5 and GD-1 data. 
This result appears compatible with results from previous
analyses using streakline models to simulate the stellar
stream orbits. \citet{bowden2015a} used a similar streakline method to model GD-1 in a logarithmic potential and found a total potential flattening for the Milky Way of $\qz=0.90^{+0.05}_{-0.09}$, and infer that the Milky Way halo is flattened at the orbital radius of GD-1. 
\citet{kupper2015a} used a streakline method to model Pal 5 in a three-component potential consisting of a variable NFW
halo and a fixed disk and bulge. 
\citet{kupper2015a} fit the main stream orbit positions,
line-of-sight velocities, and overdensities of the Pal 5 stream to the data. They found that the shape of the Milky Way halo potential is constrained to 
be $\qzh=0.95^{+0.16}_{-0.12}$, and conclude the Milky Way halo is spherical. 
We note that given their quoted uncertainties, both the \citet{bowden2015a} and \citet{kupper2015a}
analyses are consistent with the
{\it Eris} halo potential flattening ($\qzh=0.85$) to within 1$\sigma$ and permit quite oblate halos.

\subsubsection{Comparison to Previous Action-Angle Models}

Using the \cite{bovy2014a} action-angle method for
evolving the stellar stream orbits,
\citet{bovy2016a} combined the data from Pal 5, GD-1, and other Milky Way observations to constrain the shape of the Milky Way halo assuming an NFW density profile. Their Milky Way potential model consisted of a flattened dark matter halo, a Miyamoto-Nagai disk and a spherical bulge. They allowed for a variable NFW halo profile shape, leaving the 
halo density axis ratio $c/a$ as a free parameter of their model. 

When modeling Pal 5, \citet{bovy2016a} 
allowed the distance, proper motion, and
velocity dispersion of the cluster to vary. We have
adopted their priors on the Pal 5 progenitor distance and proper motions (see Table
\ref{table:priorP} for numerical values).
Given the data and priors, \citet{bovy2016a} constrained
the halo axis ratio to $c/a=0.89^{+0.20}_{-0.23}$.
They further measured the force at the location of the stream progenitor, 
and from this inferred 
that the total potential flattening was $\qz=0.94\pm0.05$.
Our {\it Eris} halo axis ratio and total potential flattening lie at the 1-$\sigma$ limit of their Pal 5 constraint. 

When modeling GD-1, \citet{bovy2016a} fixed the progenitor
cluster internal velocity dispersion to
$\sigma_v=0.4~\kms$, adopted $\phi_1=0$ assuming that the stream represents a leading arm, and varied the other five phase space coordinates of the
GD-1 progenitor as free parameters.
Following a similar analysis to Pal 5, \citet{bovy2016a} constrained the
halo density axis ratio to $c/a=1.27^{+0.27}_{-0.22}$ and the total potential flattening to $\qz=0.95\pm0.04$ for an NFW profile halo.

\citet{bovy2016a} combined the Pal 5 and GD-1 force measurements with other Milky Way potential constraints, and found a combined constraint on the axis ratio of the dark matter halo of the Milky Way of $c/a=1.05\pm0.14$ assuming the halo density profile was NFW. The dark matter halo density axis ratio for {\it Eris} ($c/a\approx0.65$) lies $\sim3\sigma$ discrepant with this result, even as {\it Eris} and a spherical NFW dark matter halo perform comparably well in reproducing the stellar stream data.
We now turn to an investigation of the origin of this apparent tension.

First, we review differences in the baryonic components of each model.
Comparing with the best fit model of \citet{bovy2016a}, {\it Eris} has a larger bulge and a smaller disk (see Table \ref{table:property} and Table \ref{table:BovyModel}).
Therefore, the contribution of the baryons to the total potential in the {\it Eris} model is more spherical and may partially compensate for the non-sphericity of the dark matter halo. To test this hypothesis, we combine the dark matter halo from {\it Eris} and the baryonic component from the spherical halo model to form a new Milky Way potential, and calculate the Bayesian evidence of this model to test
for consistency with the tidal stream data following the procedure described in Section \ref{sec:tidal streams}.
We find that this ''massive disk'' model, with a disk mass increased to
$M_d = 7.0\times10^{10}~\Msun$,
performs comparably well as the {\it Eris} or spherical
NFW halo models, with a log-evidence $\log \mathcal{Z}=-42.05\pm0.09$ for Pal 5 ($\chi^{2}=28.96$) and $\log \mathcal{Z}=-35.19\pm0.12$ for GD-1 ($\chi^{2}=18.88$), comparable to {\it Eris}.
We therefore conclude that the relative mass of the disk and bulge in {\it Eris}
does not strongly affect its ability to reproduce the stellar stream data.

The above results owe to several physical
effects. The differences in the total potential resulting
from adopting the {\it Eris} disk or the ``massive disk'' adopted from \citet{bovy2016a} are not large at the
location of the Pal 5 stellar stream. 
The potential at the location of Pal 5 is dominated by the contribution from the dark halo, and 
the influence of the disk there is weak. 
As we discussed in Section \ref{sec:shape}, the characteristic length of the disk ($\approx 2.5~\kpc$) is much smaller than the radial location of the streams. At $r \approx 15~\kpc$, the compact baryonic component acts like a point source and makes the total potential more spherical than
the dark matter potential. Therefore, 
increasing the disk mass will induce a more spherical 
total potential and using the {\it Eris} disk in our
analysis is a
conservative assumption.

We emphasize again that the disk of {\it Eris} is smaller than the Milky Way disk,
and therefore {\it Eris} does not match all constraints
on the Milky Way disk mass in the solar neighborhood. However, we do not argue that {\it Eris} is a perfect Milky Way model, but instead are
trying understand the robustness of current stellar streams on the shape of the Milky Way halo. Clearly, the (dis-)agreement between {\it Eris} and the Milky Way at the solar circle does not necessarily equate to their (dis-)agreement at 20 kpc. Replacing the {\it Eris} disk with a more massive one, as we did above, will just serve
to make {\it Eris} a ``better'' (i.e., more spherical potential) model for a Milky Way-like galaxy that can satisfy the stream constraints at both 8kpc and 20kpc.

\subsubsection{Variable Halo Shape}

We expect that the action-angle method \citep{bovy2014a} and our streakline method produce model streams of comparable physical accuracy,
and therefore our method should reproduce the
results of \citet{bovy2016a} given similar assumptions.
If not, then discrepancy between the {\it Eris} and spherical NFW halo results could
owe to the differences in modeling the streams.
We therefore revisited our analysis, allowing for an 
additional free parameter $\qzh$ 
describing the potential shape of a spheroidal NFW halo model (see Equation \ref{equ:potentialq}).
This analysis generalizes the results discussed in Section \ref{subsec:dissipation} for a spheroidal NFW
halo potential to a variable flattening $\qzh$.
We fix the baryonic component, the scale mass and the scale length of the halo to the best fit value in \citet{bovy2016a}, only allowing the potential flattening to vary within $\qzh = [0.7,1.4]$. 
Other details of the stream modeling follow the method described in Section \ref{sec:tidal streams}. 
Since both Pal 5 and GD-1 lie close to the scale length of the Milky Way halo 
\citep[$r_s=18.0\pm7.5 \kpc$;][]{bovy2016a}, we use Equation \ref{eqn:q_vs_ca} to transform the axis ratio constraints in \citet{bovy2016a} to potential shapes for a direct comparison.

We find that Pal 5 constrains the NFW potential shape to $\qzh=0.87\pm0.06$, more flattened but consistent with the value $\qzh=0.95^{+0.09}_{-0.10}$ (transformed from $c/a=0.89^{+0.20}_{-0.23}$) in \citet{bovy2016a}. {\it Eris} lies at the 1$\sigma$ limit of their result, but close to the center of our constraint. The difference in these two constraints may owe to our fixing of the baryonic galaxy parameters, while \citet{bovy2016a} allowed for a wide range of both disk and halo model parameters. Overall, we conclude our model can produce results consistent with \citet{bovy2016a} under a similar set of assumptions.

For GD-1 we find that the data prefers a prolate NFW halo with a potential flattening $\qzh=1.21\pm0.11$, in agreement with the value $\qzh=1.11^{+0.10}_{-0.09}$ (transformed from $c/a=1.27^{+0.27}_{-0.22}$) found by \citet{bovy2016a}. That the GD-1 data alone favor a prolate NFW halo (without additionally considering Pal 5) does not necessarily mean that an oblate halo will fail to reproduce the GD-1 stream data. Even though {\it Eris} has a smaller evidence as a model for GD-1 than a spherical NFW halo, we can also find a given set of initial conditions for the GD-1 progenitor 
that would allow the {\it Eris} potential to reproduce the observed streams with high fidelity (see Figure \ref{fig:gd1} and Table \ref{table:chi2}).
 
Finally, we combine the Pal 5 and GD-1 data and fit them simultaneously using an oblate NFW halo and
thirteen free parameters (four from Pal 5, eight from GD-1, and $\qzh$). We find that the NFW halo potential shape is constrained to be $\qzh = 0.97 \pm 0.07$, consistent with the results from \citet{bovy2016a}. 
This result is also consistent with a simple variance-weighted average of the constraints from Pal 5 and GD-1
separately.
We have shown that the oblate {\it Eris} halo is consistent with the data, but we agree with \citet{bovy2016a} that for an NFW profile the 
data favor a spherical halo. 
Since we have used the same stream modeling method and the same parameter setup, the only difference between these two analyses is the form of the potential, i.e., {\it Eris} differs from an analytical oblate NFW halo. {\it Eris} is a dark halo formed self-consistently in a cosmological simulation that models the dissipative
physics relevant for the formation of a realistic Milky Way-like galaxy. Given the comparable quality of the
fits and the current constraining power of the data, we suggest that both the oblate {\it Eris} halo and a spherical NFW halo provide reasonable models
for dark matter halos in galaxies like the Milky Way.


\subsubsection{Comparison to Other Tidal Stream Analyses}

While we use the {\it Eris} hydrodynamical simulation to provide
the background potential for our stellar stream modeling,
prior work has used collisionless dark matter-only simulations
to model the development of Milky Way stellar streams.
\citet{ngan2015a} and \citet{ngan2016a} simulated tidal streams in the potential of the Via Lactea \uppercase\expandafter{\romannumeral2} (VL2) N-body
simulation and a spherical analytical potential, with and without subhalos, to study the stream morphologies. They used a similar SCF approach to ours for representing
the simulated dark matter halo,
except they used the \citet{hernquist1990a} radial basis functions instead of
the NFW model we adopt. \citet{ngan2015a} and \citet{ngan2016a}  found that
orbiting subhalos and the time-dependence of the potential shape could
lead to both fluctuations in the density and eventually the dispersal of the
tidal streams. They also found that massive subhalos could increase the stream
density by shepherding stream stars.

More recently, \citet{sandford2017a} used the VL2 potential and other, more
idealized smooth potentials to examine further the interplay between
stream morphologies, the global halo shape, and the presence of substructures.
\citet{sandford2017a} found that the shape of the halo potential and substructure
could
affect stream morphological properties such as their thickness or
the regularity of their stellar distribution. 

Our work largely complements these previous efforts, in that we only use the positions
and velocities of the streams to evaluate {\it Eris} as a model of a Milky Way-like
halo and do not attempt to use the 
morphological properties of the streams to evaluate the model potential. We do
find that the aspherical halo in the dark matter-only {\it ErisDark} simulation
widens the model streams. The end of the Pal 5 trailing arm becomes relatively broad, even though we set the internal velocity dispersion of the progenitor cluster to $\sigma_v=0$, and the test particles take twice as long cover the whole Pal 5 stream in
{\it ErisDark} than in {\it Eris}. With a non-zero
progenitor internal velocity dispersion the stream-fanning in {\it ErisDark} would be exacerbated \citep{pearson2015a}. The same phenomenon also occurs in the GD-1 model.

\begin{figure*}[htp]
\centering
\includegraphics[width=\linewidth, clip]{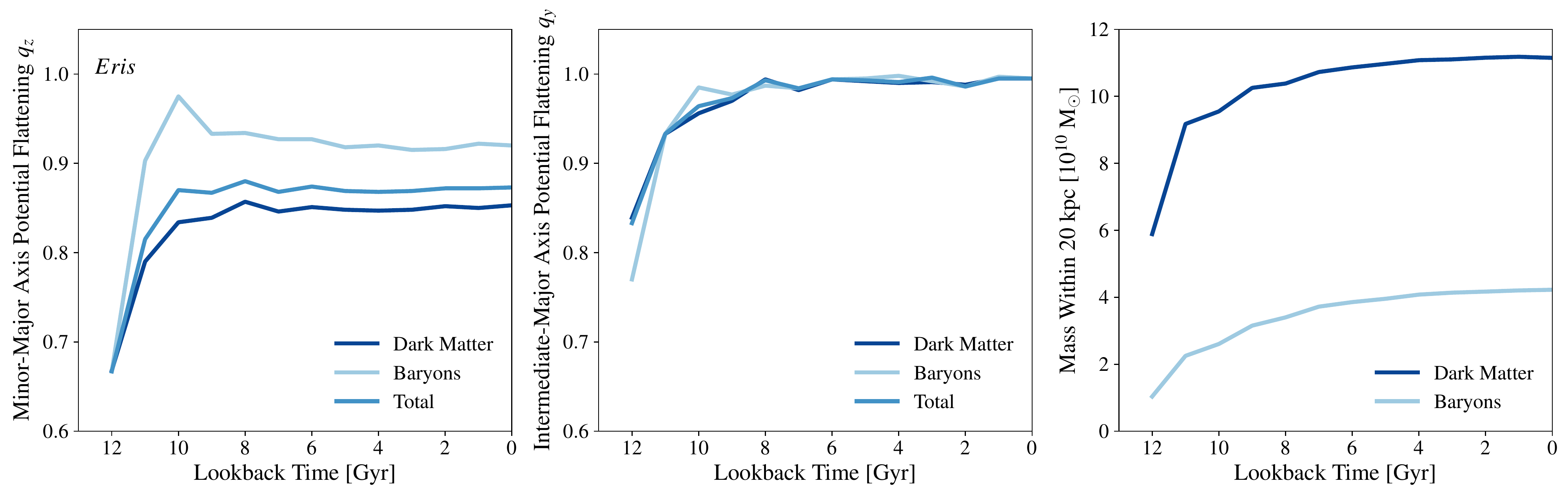}
\caption{Time evolution of the {\it Eris} dark matter halo. The left and middle panels show the minor-major and intermediate-major axis ratios of the {\it Eris} halo potential as a function of time, split by the dark matter (dark blue), baryons (light blue), and total potential (medium blue) components. The axis ratios are measured at $r=15~\kpc$, close to where Pal 5 and GD-1 are located. The right panel displays the evolution of mass within $20~\kpc$ of the halo for the dark matter (dark blue) and baryons (light blue). These results suggest that the halo remains stable over the last $\sim4\Gyr$.}
\label{fig:history}
\bigskip
\end{figure*}

\subsection{{\it Gaia} and Future Observations}

Given the current quality of data, we have demonstrated that both {\it Eris} and a spherical halo model prove consistent with the current Pal 5 and GD-1 stream data, even though the models differ structurally. Relative to our spherical halo model,
{\it Eris} has a smaller disk, less mass within 20kpc,
and a substantially flattened halo ($\qzh=0.85$), but both models perform
comparably well. However, with more precise data one or both models might
be falsifiable. Based on our fitting to the Pal 5 stellar stream, we constrain the Pal 5 globular cluster parameters to be [$D$, $\mu_{\alpha}$, $\mu_{\delta}$, $M$]$=$[$21.0\pm 0.8~\kpc$,
$-2.337\pm 0.096~\masyr$, $-2.302\pm 0.100~\masyr$, 
$1.3\pm 0.6\ \times10^4~\Msun$] in {\it Eris} and [$21.7\pm 0.5~\kpc$, $-2.393\pm 0.058$ $\masyr$, $-2.349\pm 0.061$ $\masyr$, $1.4\pm 0.4 \times10^4~\Msun$] in the spherical halo model. We find that proper motion is the most sensitive parameter for distinguishing between different potentials, as further indicated by the one dimensional marginalized posterior distributions shown in Figure \ref{fig:Pal5posterior}.
Based on the posterior distribution from the {\it Eris} model, we estimate that improving the proper motion constraints to $\sigma_{\mu}\lesssim50~\muasyr$ for the
streams would distinguish between {\it Eris} and a more
spherical halo, well beyond the current $\sigma_{\mu}\approx 150~\muasyr$ uncertainties.
According to \citet{de_bruijne2012a},
{\it Gaia} should reach final mission
proper motion constraints
of $\sigma_{\mu}\lesssim100~\muasyr$ for individual stars
with the apparent brightness of stars in the Pal 5 and
GD-1 streams ($m\approx17-20$), and will provide much
tighter constraints by combining measurements from
multiple stars in each stream. We therefore expect
the final mission {\it Gaia} data will be able to distinguish between models
like {\it Eris} and a more spherical halo model for the Milky
Way.

\subsection{Time Evolution of the {\it Eris} Halo}
\label{subsec:evolution}

Dark matter halos evolve over time owing to accretion. If the shape of the host halo potential changes dramatically during the period of the formation of a tidal stream, using a static halo potential to model the tidal stream could be inappropriate. We modeled the potential of {\it Eris} halo at redshift $z=0$ using our basis expansion, and we assumed that the halo did not evolve over the
formation time of the stellar streams. To test the accuracy of this assumption, we measure the shape and inner mass at $r\lesssim20~\kpc$ of the {\it Eris} halo as a function of time, as shown in Figure \ref{fig:history}. The shape of the {\it Eris} halo evolves very slowly over the last 8 billion years. Over the same
period, the inner halo increases its mass by only about 10\% and remains quite
stable over the last 4 billion years. In our model the Pal 5 and GD-1 
stellar streams develop
over the last $\sim4$ billion years, and these measurements validate our
use of a static halo when using Pal 5 and GD-1 streams to analyze the
constraints on the Milky Way potential.

\subsection{Caveats}
\label{subsec:caveats}
The results of our analysis are subject to important caveats.
The comparison of simulated galaxies like {\it Eris} with real systems like the Milky Way often involve
an imperfect statistical rationale, as the expense of high-resolution cosmological simulations 
typically limit the range of model galaxies available to compare with observations. Our analysis
deals with only a single realistic Milky Way analog, and it remains unclear whether the results
we present for {\it Eris} generalize to all simulated galaxies that resemble the Milky Way. We
plan to revisit our results using the {\it Venus} simulation \citep[see][]{sokolowska2017a}, a
similar simulation to {\it Eris} but where the resimulated galaxy is chosen to have a more violent
merger history. 
Even given these caveats, our results do indicate that both the halo mass distribution and its flattening 
matter in shaping stellar streams, and appropriate caution should be employed when assuming a
halo density profile while attempting to infer the flattening of the halo potential.

\section{CONCLUSIONS}
\label{sec:conclusions}

We have used the {\it Eris} hydrodynamical simulation of a Milky Way-like
galaxy forming in a full cosmological context \citep{guedes2011a} as a model for reproducing the
observed tidal streams Pal 5 and GD-1. We find that combined
baryonic and dark matter components of {\it Eris} produce a total gravitational
potential shape of $\qz = 0.88$, and that this shape allows for the formation
of tidal streams that show remarkable consistency with the Pal 5 and GD-1 data. 
Using Bayesian inference we demonstrate that
the {\it Eris} model works as well as the
best fit (spherical) potential presented by \citet{bovy2016a}.
Remarkably,
the tidal streams Pal 5 and GD-1 can be well-reproduced by the potential of {\it Eris} without any rescaling or modifications,
suggesting that our current models of
$\LCDM$ structure formation including dissipative physics
can produce galaxies that have total halo potentials in substantial agreement
with current observational constraints available for the Milky Way.

We find that
the total potential {\it Eris} works comparably well to the spherical NFW halo model
found by \citet{bovy2016a}, even
though the dark matter halo in {\it Eris} has an axis ratio of $c/a=0.65$. 
The baryonic dissipation alters the dark matter halo shape from 
prolate to an oblate halo aligned with the disk, consistent with
previous studies \citep[e.g.,][]{kazantzidis2004a, gustafsson2006a, abadi2010a}. 
The more centrally-concentrated mass distribution of {\it Eris} relative to 
an NFW halo allows for {\it Eris} to permit realistic stream orbits, despite its flattened shape.
The {\it Eris} halo
shape is substantially rounder than in the equivalent dark matter-only simulation
{\it ErisDark}; we find that {\it ErisDark} cannot reproduce the observed stellar
stream data.

With the on-going release of the {\it Gaia} data, 
we expect exciting additional probes of the Milky Way potential that
will further test our models for $\LCDM$ galaxy formation.
{\it Gaia} will provide a full-mission proper motion
precision of $\sigma_{\mu}\lesssim100~\muasyr$ for single stars
down to 20$^\mathrm{th}$ magnitude \citep{de_bruijne2012a}.
With multiple available stars at magnitudes
$m<20$ in each stream,
the constraints on both the Pal 5 and GD-1 
proper motions should sharpen substantially
beyond the $\sigma_{\mu}\approx150~\muasyr$
uncertainties from the current data \citep{koposov2010a,fritz2015a}. The ultimate
precision afforded by {\it Gaia} should
distinguish between models like {\it Eris} and more
spherical halos, if the final stream proper motion 
constraints reach $\sigma_{\mu}\lesssim50~\muasyr$.

\acknowledgments

We thank the anonymous referee for comments that helped improve our work.
We also thank Annalisa Pillepich for providing access to the {\it ErisDark} simulation. Support for this work was provided by the Chinese Scholarship Council (BD), and by NASA through grant HST-AR-13904.001-A (PM) and contract NNG16PJ25C (BER). PM also thanks the Pr\'{e}fecture of the Ile-de-France Region for the award of a Blaise Pascal International Research Chair, managed by the Fondation de l'Ecole Normale Sup\'{e}rieure. This research utilized the NSF-funded Hyades supercomputer at UCSC.

\bibliography{main}

\end{document}